# A Low Temperature Kinetic Study of the C($^3$P) + CH$_3$OCH$_3$ Reaction. Rate constants, H-atom Product Yields and Astrochemical Implications

Kevin M. Hickson,[1,*] Jean-Christophe Loison,[1] and Valentine Wakelam[2]

[1]Institut des Sciences Moléculaires ISM, CNRS UMR 5255, Univ. Bordeaux, 351 Cours de la Libération, F-33400, Talence, France

[2]Laboratoire d'astrophysique de Bordeaux, CNRS, Univ. Bordeaux, B18N, allée Geoffroy Saint-Hilaire, F-33615 Pessac, France

**Abstract**

Atomic carbon in its ground electronic state, C($^3$P), is expected to be present at high abundances during the evolution of dense molecular clouds. Consequently, its reactions with other interstellar species could have a strong influence on the chemical composition of these regions. Here, we report the results of an investigation of the reaction between C($^3$P) and dimethylether, CH$_3$OCH$_3$, which was recently detected in dark cloud TMC-1. Experiments were performed to study the kinetics of this reaction using a continuous supersonic flow reactor employing pulsed laser photolysis and pulsed laser induced fluorescence for atomic radical generation and detection respectively. Rate constants for this process were measured between 50 K and 296 K, while additional measurements of the product atomic hydrogen yields were also performed over the 75-296 K range. To better understand the experimental results, statistical rate theory was used to calculate rate constants over the same temperature range and to provide insight on the major product channels. These simulations, based on quantum chemical calculations of the ground triplet state of the C$_3$H$_6$O molecule, allowed us to obtain the most important features of the underlying potential energy surface. The measured rate constant increases as the temperature falls, reaching a value of $k_{C+CH_3OCH_3} =$ 7.5 × 10$^{-11}$ cm$^3$ s$^{-1}$ at 50 K, while the low measured H-atom yields support the theoretical prediction that the major reaction products are CH$_3$ + CH$_3$ + CO. The effects of this reaction on the abundances of interstellar CH$_3$OCH$_3$ and related species were tested using a gas-grain model of dense interstellar clouds, employing an expression for the rate constant, k(*T*) = α(*T*/300)$^\beta$, with α = 1.27 × 10$^{-11}$ and β = -1.01. These simulations predict that the C($^3$P) + CH$_3$OCH$_3$ reaction decreases gas-phase CH$_3$OCH$_3$ abundances by more than an order of magnitude at early and intermediate cloud ages.

**Introduction**

Carbon is the fourth most abundant element after hydrogen, helium and oxygen and is present throughout the Universe as an isolated species, C in both its neutral and ionic forms, as well as being incorporated into numerous organic type molecules. In the interstellar medium (ISM), the gravitational collapse of diffuse clouds of gas and dust forming denser, darker regions is accompanied by a transformation of the underlying composition. This occurs due to a change in the physical conditions, from regions where photons dominate so that the chemistry involves atoms and ions, to regions where photons play a much less important role and the chemistry is dominated by reactions between neutral species. Under these conditions, elemental carbon initially present as $C^+$ due to its low ionization potential is rapidly converted to neutral atomic carbon in its ground electronic state, $C(^3P)$. Indeed, observations of $C(^3P)$ atoms towards dense molecular clouds such as TMC-1,[1] OMC-1[2,3] and Barnard 5[4] through its fine structure transitions in the submillimeter-wave range (492.162 and 809.345 GHz) clearly indicate that neutral atomic carbon is present within these clouds at high abundance levels, coexisting alongside major reservoir species of C and O such as CO. As astrochemical models[5] predict that neutral atomic carbon abundances are high ($10^{-4}$ with respect to $H_2$) during the early stages of cloud evolution, it is important to evaluate the importance of the reactions of $C(^3P)$ atoms with other interstellar species to examine their effects on the overall composition of these regions. Indeed, the dense ISM is home to a wide variety of organic molecules ranging from simple diatomic radicals such as CH to large species such as $C_{70}$.[6] While there are many previous studies of the reactivity of $C(^3P)$ with unsaturated hydrocarbons[7-16], there are relatively few studies of its reactions with organic molecules containing other functional groups[17,18] such as those found in complex organic molecules (COMs). COMs, arbitrarily defined as carbon bearing molecules containing six or more atoms[19] have been detected in a wide variety of interstellar environments ranging from warmer regions such as protostellar envelopes,[20] hot cores[21] and hot corinos[22] to cold objects such as prestellar cores.[23] Among these COMs, dimethyl ether, $CH_3OCH_3$, was first detected towards the Orion molecular cloud in 1974 by Snyder et al.[24] Although it has since been observed in numerous "warm" interstellar objects including low,[25] intermediate[26] and high mass star forming regions,[27] it is not restricted to these environments where ice sublimation is expected to play an important role by boosting gas-phase COM abundances. Indeed, $CH_3OCH_3$ has been detected recently[28] in the prototypical dark cloud TMC-1 with an abundance of $2.5 \times 10^{-10}$ with respect to $H_2$ and towards the prestellar core L1689B[23] with a similar abundance of a few $10^{-10}$. In astrochemical databases such as the Kinetic Database for Astrochemistry (KIDA),[29] $CH_3OCH_3$ formation in the gas-phase occurs through only one dissociative electron recombination reaction at the present time, namely the $CH_3OCH_4^+ + e^- \rightarrow CH_3OCH_3 + H$ reaction. The radiative association reaction $CH_3O + CH_3 \rightarrow CH_3OCH_3$ + photon suggested by Balucani et al.[30] and the $(CH_3)_2OH^+ + NH_3 \rightarrow CH_3OCH_3 + NH_4^+$ ion molecule reaction proposed by Skouteris et al.[31] are not yet included in KIDA. Instead, $CH_3OCH_3$ is considered to be synthesized primarily on interstellar ices, by a range of processes including the sequential hydrogenation of simpler carbon and oxygen bearing molecules such as $CH_3OH$.[32] Despite this, models typically underestimate the observed gas-phase $CH_3OCH_3$ abundances in the dense ISM either due to the difficulty of reinjecting these molecules into the gas-phase in sufficient quantities at such low temperatures (10 K or lower) or because important gas-phase or surface reactions are missing.

To improve our understanding of the interstellar chemistry of dimethyl ether, we performed an experimental and theoretical study of the $C(^3P) + CH_3OCH_3$ reaction. On the experimental side, a supersonic flow apparatus has been employed to investigate the kinetics of this reaction over the 50-296 K range, coupled with pulsed laser photolysis and pulsed laser induced fluorescence for the production and detection of $C(^3P)$ atoms in the cold supersonic flow. In addition to the kinetic studies, measuring temperature dependent rate constants for this process, we have also measured those product channels leading to H-atom formation at room temperature and below. In conjunction with electronic structure calculations of the underlying potential energy surface, and a Rice Ramsperger Kassel Marcus (RRKM) master equation (ME) analysis of the major pathways, it has been possible to identify the major product channels of this reaction for astrochemical modeling purposes. Finally, the effects of the $C(^3P) + CH_3OCH_3$ reaction on interstellar chemistry have been tested through an astrochemical modeling study employing both gas-phase and grain surface reactions.

The paper is organized as follows. The experimental and theoretical methods are described in sections 2 and 3 respectively, while the results of this work are presented in section 4. The astrochemical simulations and the implications of this study for interstellar dimethyl ether and related species are given in section 5, followed by our conclusions in section 6.

## 2 Experimental Methods

An existing continuous supersonic flow reactor was used to perform the experimental work described here.[33, 34] Subsequent modifications to the apparatus, particularly on the detection side, have allowed us to perform kinetic studies of the reactions of atomic radicals in both the ground ($H(^2S)$,[35] $C(^3P)$,[36] $N(^4S)$[37]) and excited ($O(^1D)$,[38] $N(^2D)$[39]) electronic states. Here, convergent-divergent Laval type nozzles[40] were employed to produce supersonic flows of a specified carrier gas (Ar or $N_2$ in this instance) with the uniform temperature, density and velocity profiles required to study the kinetics of the $C(^3P) + CH_3OCH_3$ reaction below room temperature. As these flows typically persist for only a few tens of centimetres, such methods are generally applicable to the study of fast reactions only (i.e., with rate constants > $10^{-12}$ cm$^3$ s$^{-1}$). Three different nozzles were employed during this investigation to access four different low temperatures (one nozzle was used with both Ar and $N_2$), allowing flows with characteristic temperatures of 177 K, 127 K, 75 K and 50 K to be generated. The flow properties for each nozzle are summarized in Table 2 of Hickson et al.[36] The nozzle was mounted on a piston that could slide in and out of the reactor, allowing it to be positioned at a specified distance from the observation axis. For any particular nozzle this value corresponded to the maximum distance for which the flow conditions were considered to remain optimal. These characteristic distances were derived during earlier calibration experiments examining the variation of the supersonic flow impact pressure as a function of distance from a Pitot tube.[33] To perform experiments at room temperature (296 K), the nozzle was removed, while the flow velocity was decreased to eliminate pressure gradients within the chamber.

$C(^3P)$ atoms were generated in situ during these experiments by the 266 nm pulsed (10 Hz) photolysis of trace amounts of tetrabromomethane ($CBr_4$) molecules in the supersonic flow. To carry these molecules into the reactor, a small fraction of the carrier gas was diverted into a vessel at room temperature containing $CBr_4$ crystals before rejoining the main flow upstream of the Laval nozzle reservoir. For any series of measurements at a given flow temperature, the pressure in the vessel was set to a fixed value by adjusting a needle valve at the outlet. Based on its saturated vapour pressure at room temperature (0.7 Torr), the $CBr_4$ concentration in the cold flow was estimated to be less than $2.6 \times 10^{13}$ cm$^{-3}$. Pulse energies of approximately of 30-38 mJ with a beam diameter of 5mm were used here. The photolysis beam itself was coaligned along the supersonic flow by sending it through a quartz window at the Brewster angle at the back of the reactor. The beam exited the reactor through the nozzle throat and a second Brewster angled quartz window attached to the back of the reservoir. As such, C($^3$P) atoms were generated along the supersonic flow with identical concentrations at any axial position as the UV beam is only weakly attenuated by CBr$_4$ ($\sigma_{CBr_4}(266nm) = 1 \times 10^{-18}$ cm$^2$).

C($^3$P) atoms were detected by pulsed (10 Hz) laser induced fluorescence at vacuum ultraviolet wavelengths (VUV LIF) via the 2s$^2$2p$^2$ $^3$P$_2$ → 2s$^2$ 2p5d $^3$D$_3$° transition at 115.803 nm. To generate this wavelength, we first frequency doubled the output of a Nd:YAG pumped dye laser at 695 nm. The residual 695 nm dye laser radiation was then discarded using two dichroic mirrors coated for peak reflectivity at 355 nm so that only the 347 nm UV beam remained. Some of the residual dye laser radiation was fed into the output coupler of a wavemeter to allow us to follow the probe laser wavelength continuously throughout the experiment. The 347 nm beam was directed towards the observation axis of the reactor by right angled quartz prisms where it was focused into a gas cell containing xenon (50 Torr) with argon (160 Torr) added for phase-matching to generate VUV radiation by frequency tripling. H($^2$S) atoms were also detected during this work, as products of the C + CH$_3$OCH$_3$ reaction. This was done using a similar procedure to the one described above for C($^3$P) atoms to generate a tunable VUV beam around the Lyman-α line at 121.567 nm starting from narrow-band dye laser radiation at 729.4 nm. In this case, krypton (210 Torr) was used as the tripling medium with argon (540 Torr) added for phase-matching. A MgF$_2$ lens was used as the exit window of the cell to recollimate the VUV beam, leaving the residual UV beam divergent due to the difference in refractive index of MgF$_2$ at UV and VUV wavelengths. As the cell was attached to the reactor at right angles via a 75 cm long sidearm containing circular baffles, only a low flux of UV radiation reached the observation region. The sidearm was open to the reactor, containing the same residual gases as those present in the reactor including CBr$_4$ and CH$_3$OCH$_3$ which both absorb strongly in the VUV region. As significant attenuation of the VUV probe beam was likely to occur, this region was continuously flushed with either argon or nitrogen during the experiments, thereby preventing residual gases from filling the sidearm. Additional test experiments were performed to check for undesirable effects brought about by the residual UV radiation or the generated VUV radiation. To test for the influence of the residual UV beam entering the reactor, the tripling cell was evacuated to ensure that no C($^3$P) and H($^2$S) signals were observed in the absence of VUV light. To check for the influence of the VUV beam alone, experiments were performed with the photolysis laser off (with gas in the tripling cell) to check for the formation of C($^3$P) and/or H($^2$S) by the VUV beam. No signal from C($^3$P) was ever observed, but a very small H-atom fluorescence signal was observed with CH$_3$OCH$_3$ present in the flow, due to CH$_3$OCH$_3$ photolysis by the VUV probe beam at 121.567 nm. As this small fluorescence signal represented the same constant contribution for all time points during a single run, this was automatically removed by the baseline subtraction procedure. Resonant emission from the C($^3$P) or H($^2$S) atoms in the cold supersonic flow was detected at right angles to both the supersonic flow and the VUV probe laser to minimize scattered light detection from both the photolysis laser and the probe laser. The fluorescence was focused by a lithium fluoride (LiF) lens onto the photocathode of a solar blind photomultiplier tube (PMT – Electron Tubes 9403B). Both the LiF lens and the PMT were isolated from the chamber by a LiF window, thereby avoiding damage to the PMT input window by reactive gases. Atmospheric absorption losses of the VUV fluorescence were eliminated by evacuating the zone containing the LiF lens using a dry vacuum pump. The PMT output signal was processed by a boxcar integration system, while the lasers, boxcar and oscilloscope were synchronized by a digital delay generator operating at 10 Hz. For the present experiments, the C- or H-atom fluorescence signal was measured as a function of delay time between the two lasers. 30 laser shots were recorded and averaged for each time point, with at least 100 points recorded for each temporal trace. 15 of these time points were recorded with the probe laser firing prior to the photolysis laser so unwanted contributions to the signal such as electronic noise, scattered light and the H-atom fluorescence signal generated by CH$_3$OCH$_3$ photolysis by the probe laser could be subtracted.

## 3 Electronic Structure Calculations

The C($^3$P) + CH$_3$OCH$_3$ reaction occurs over the ground state $^3$A' potential energy surface of C$_3$H$_6$O assuming C$_1$ geometry. The geometries and energies of the reagents/products, intermediates and transition state structures (TSs) were calculated initially using density functional theory (DFT), specifically using the hybrid meta exchange correlation functional, M06-2X,[41] coupled with the aug-cc-pVTZ (AVTZ) basis set (M06-2X/AVTZ). The calculation of harmonic frequencies for these structures allowed us to determine the nature of the stationary point (no imaginary frequencies for a reagent/product species or minimum, a single imaginary frequency for a TS) and obtain the zero-point energy (ZPE) corrections for all structures. Then, for each identified TS, an intrinsic reaction coordinate calculation was performed to ensure that it lies on the minimum energy pathway between two specified minima. Then, as a second step, we calculated the energies of all species more accurately using domain based local pair-natural orbital singles and doubles coupled cluster theory with an improved perturbative triples correction algorithm (DLPNO-CCSD(T))[42] as integrated in the ORCA computational package,[43,44] adopting the geometries derived at the M06-2X/AVTZ level. The DLPNO-CCSD(T) calculations were coupled with the AVTZ basis set here (DLPNO-CCSD(T)/AVTZ). Although all of the underlying calculations were performed with ORCA, the structures were visualized and manipulated using Avogadro,[45] which was also used to perform the vibrational frequency analysis.

## 4 Results and discussion

### 4.1 Potential Energy Surface

The calculated ground state triplet surface ($^3A'$) involved in the C + $CH_3OCH_3$ reaction is shown in Figure 1.

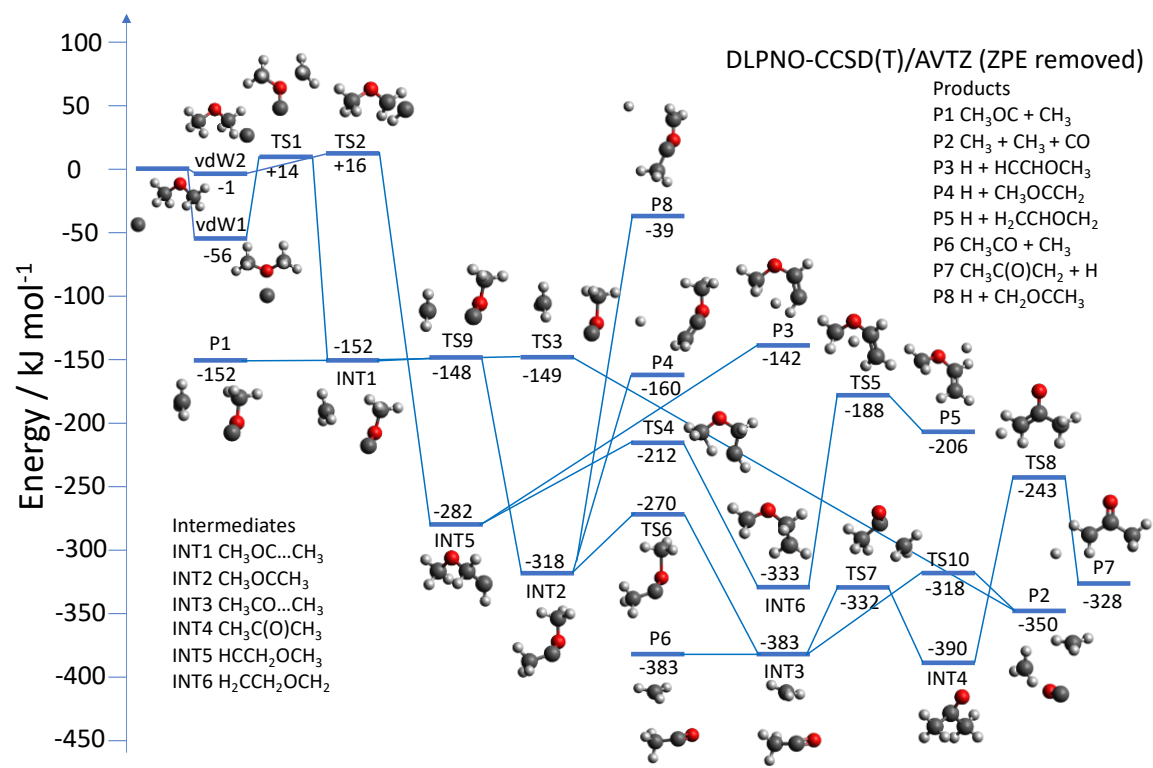

**Figure 1** Ground state triplet potential energy surface for the C($^3$P) + CH$_3$OCH$_3$ reaction. All energies are calculated at the DLPNO-CCSD(T)/AVTZ level based on geometries optimized at the M06-2X/AVTZ level. Displayed energies are relative to the reagent C($^3$P) + CH$_3$OCH$_3$ asymptote and are corrected for ZPE differences.

All energies are quoted at the DLPNO-CCSD(T)/AVTZ level relative to the C + CH$_3$OCH$_3$ entrance channel and are corrected for ZPE differences unless otherwise stated. 8 distinct minima were identified, connected by 10 transition states, leading to 8 different exit channels. The interaction of ground triplet state carbon with CH$_3$OCH$_3$ leads to the formation of two possible van der Waals complexes.

Initial carbon attack at the central oxygen atom of CH$_3$OCH$_3$ leads to the formation of vdW1 (-56 kJ/mol) with a distance of 1.645 Å between O and the incoming C-atom. vdW1 can then evolve further via TS1 +14 kJ/mol above the reagent level corresponding to the formation of a chemical bond between the incoming C and O followed by breaking of one of the O-CH$_3$ bonds of CH$_3$OCH$_3$. The resulting weakly bound CH$_3$OC...CH$_3$ complex (INT1) is calculated to be -152 kJ/mol below the reagent level. Three possible channels are open to INT1. (I) it can dissociate to form CH$_3$OC + CH$_3$ products (P1) also calculated to be -152 kJ/mol below the reagents. It should be noted here that although the final ZPE corrected DLPNO-CCSD(T) energies are essentially identical, both M06-2X/AVTZ and DLPNO-CCSD(T)/AVTZ calculations indicate that INT1 is more stable than P1 products by -12 and -4 kJ/mol respectively if ZPE differences are not considered. (II) INT1 can isomerize to form the CH$_3$OCCH$_3$ intermediate INT2 at -318 kJ/mol, over a very low barrier, TS9, only 4 kJ/mol above INT1. (III) INT1 can dissociate further to products P2, CH$_3$ + CH$_3$ + CO (-350 kJ/mol) over another very low barrier, TS3, only 3 kJ/mol above INT1. Of the three possible channels, only INT2 formed by isomerization channel (II) will evolve further. One possibility is that of C-H bond dissociation leading to either H + CH$_2$OCCH$_3$ products P8, -39 kJ/mol below the reagent level or H + CH$_3$OCCH$_2$ products (P4), -160 kJ/mol below the reagent level. While scans along the reaction coordinate at the M06-2X/AVTZ level confirmed that the INT2→P4 channel presents no barrier to dissociation, it was not possible to extract any reliable information regarding the INT2→P8 channel at the M06-2X/AVTZ level due to the multiconfigurational nature of this bond dissociation process. Instead, complete active space self-consistent field (CASSCF) calculations using six active orbitals and six active electrons were performed with the aug-cc-pVDZ basis set, followed by Davidson corrected multireference configuration calculations (MRCI+Q) with the same active space to obtain more accurate energies. These additional calculations were also performed using ORCA.[43] During these calculations, the molecule was optimized for each value of the dissociating bond distance between 1 and 4 Å at the CASSCF level while a vibrational frequency analysis was also performed at each step. The energies and vibration frequencies derived in this way were used as inputs in subsequent kinetic calculations described in section 4.2.

Another possibility for evolution of the INT2 intermediate is for the remaining C-O bond of the original $CH_3OCH_3$ molecule (the $H_3C-OCCH_3$ bond) to dissociate to form the intermediate $CH_3CO…CH_3$ species INT3 (-383 kJ/mol) over TS6, 48 kJ/mol above INT2. In a similar manner to INT1 described above, INT3 can (I) dissociate to products P6, $CH_3CO + CH_3$ (-383 kJ/mol), (II) dissociate further to P2 (-350 kJ/mol) over TS10, 65 kJ/mol above INT3, (III) isomerize to INT4, $CH_3C(O)CH_3$ (-390 kJ/mol), over TS7, 51 kJ/mol above the INT3 level. INT4 can then lose a hydrogen atom to form products P7, H + $CH_3C(O)CH_2$ (-328 kJ/mol), by passing over TS8, 147 kJ/mol above INT4.

In an alternative pathway, carbon attack can initially occur at one of the terminal hydrogen atom positions of one of the methyl groups of $CH_3OCH_3$, leading to the very weakly bound complex vdW2 (-1 kJ/mol) with a newly formed C-H bond distance of 1.500 Å. Given the very low energy of this complex, which is well within the expected theoretical uncertainty of the DLPNO CCSD(T) method (larger than the 4 kJ/mol chemical accuracy of canonical CCSD(T)), this pathway is not considered further in the subsequent kinetic calculations. Nevertheless, for completeness we describe the potential pathways originating from this complex. The weakly bound C-atom can insert into the C-H bond of the methyl group over TS2, +16 kJ/mol above the reagent level, leading to the formation of INT5, $HCCH_2OCH_3$ (-282 kJ/mol). INT5 can then evolve further by losing a hydrogen atom leading to products P3, H + $HCCHOCH_3$ (-142 kJ/mol) (only one dissociation pathway is shown here), or by isomerization through H-atom transfer to INT6, $H_2CCH_2OCH_2$ (-333 kJ/mol). INT6 can dissociate by losing a hydrogen atom through several pathways such as the one over TS5 to form products P5 H + $H_2CCHOCH_2$ (-206 kJ/mol). Other pathways, such as the abstraction pathway leading to CH + $CH_2OCH_3$ formation were not considered here as these products are calculated to be endothermic with respect to the C + $CH_3OCH_3$ entrance channel by +65 kJ/mol at the DLPNO CCSD(T)/AVTZ level.

It should be noted that neither of the two reaction pathways presented here (leading from C + CH$_3$OCH$_3$ → vdW1 or vdW2 → products) were found to be barrierless at the DLPNO-CCSD(T)/AVTZ//M06-2X/AVTZ level and no other potentially important barrierless reaction channels were identified. Consequently, if we consider the nominal values of the TS1 and TS2 barriers on the ground state PES then the C + CH$_3$OCH$_3$ reaction is expected to be slow below room temperature in the absence of an important tunneling contribution. The possibility of intersystem crossing to the singlet state, in a similar manner to a recent study of the O($^3$P) + pyridine reaction,[46] was also considered in the present work. MRCI+Q calculations (2e$^-$, 4o) based on CASSCF wavefunctions performed on the C…O(CH$_3$)$_2$ entrance channel showed that the singlet surface (correlating with C($^1$D) + CH$_3$OCH$_3$ reagents) crosses the triplet surface in the vicinity of TS1, with an energy slightly above the C($^3$P) + CH$_3$OCH$_3$ asymptote. As the CASSCF wavefunctions lead to monoconfigurational wavefunctions, DLPNO-CCSD(T)/AVTZ//M06-2X/AVTZ calculations of the minimum energy crossing point (MECP) should be valid, leading to an energy of +39 kJ/mol for the MECP, even higher than the TS1 and TS2 energies. The geometries and frequencies of all the structures presented in Figure 1 are listed in the supplementary information file.

**4.2 Kinetic calculations**

According to the electronic structure calculations presented above, the C + CH$_3$OCH$_3$ reaction is expected to occur through the initial barrierless formation of a complex (vdW1, assuming that vdW2 is too weakly bound to play any significant role in the reaction) followed by isomerization of this chemically activated species to form the CH$_3$OC…CH$_3$ (INT1) intermediate. Under these conditions, the predicted overall reaction rate at a specific temperature can be derived from the calculated capture rate constant $k_{capt}(T)$ to form vdW1 from the separated reagents, multiplied by the fraction of the population in vdW1 that reaches INT1 by passing over TS1. The rate constant calculations reported here were performed using the *Multiwell* suite of programs[47, 48] in a similar manner to our earlier study of the C + N$_2$O reaction.[5] In addition, we have extended the scope of the present work by calculating the product branching fractions at selected temperatures, allowing us to compare directly with the measured H-atom yields.

The initial step was to calculate the energy ($E$) and angular momentum ($J$) resolved microcanonical rate constants $k(E,J)$ for the unimolecular dissociation of vdW1 to C +

CH₃OCH₃ using the *ktools* program[47] which implements a variational treatment due to the lack of a clearly defined barrier for this process. In order to apply this method, it was necessary to perform a series of constrained optimizations along the reaction coordinate from the equilibrium C..O bond distance of 1.645 Å in vdW1 to the separated reagents. These optimizations were performed at the M06-2X/AVTZ level by increasing the C-O bond distance by 0.2 Å steps along the reaction coordinate until the separation between the carbon and oxygen atom of CH₃OCH₃ reached 6 Å. A subsequent vibrational frequency calculation was then performed at the same level for all of these optimized structures. The harmonic vibrational frequencies orthogonal to the reaction path, the calculated rotational constants and the energies of these structures in addition to those of the separated reagents and vdW1 complex (all calculated at the DLPNO-CCSD(T)/AVTZ level) served as inputs for the *ktools* calculations[47] which were performed using a small grain size of 1 cm⁻¹ for better accuracy at low temperature. The calculated frequencies and energies are listed in Table S4. Trial rate constants were calculated for each fixed distance, with the variational transition state given by the point yielding the minimum trial rate constant value. When multiple significant minima were located, as was the case for the entrance channel leading to vdW1, a multiple transition state model was applied based on the unified statistical theory of Miller.[49]

Averaging the values of $k(E,J)$ over *E* and *J* at a given temperature,[50] allowed us to calculate the canonical rate constants, equivalent to the high-pressure-limit rate constant, $k_{\infty,}^{uni}(T)$. The capture rate constant, $k_{capt}(T)$, was then determined from the equilibrium constant, $K_{eq}(T)$, through expression (1)

$$k_{capt}(T) = k_{\infty,}^{uni}(T)/K_{eq}(T) \qquad (1)$$

The values of $k_{capt}(T)$ derived in this way for a range of temperatures between 50 and 296 K were used as inputs for subsequent master equation (ME) simulations to obtain the final overall temperature and pressure-dependent rate constants and product branching ratios. RRKM statistical theory as implemented in *Multiwell*[47] was employed to derive energy-dependent microcanonical rate constants k(*E*) for a specific reaction step

$$k(E) = L^{\neq} \frac{g_e^{\neq}}{g_e} \frac{1}{h} \frac{G^{\neq}(E-E_0)}{\rho(E)} \qquad (2)$$

where $L^{\neq} = \frac{m^{\neq}}{m} \times \frac{\sigma_{ext}}{\sigma_{ext}^{\neq}}$ is the reaction path degeneracy, with $m^{\neq}$ and $m$ the number of optical isomers of the TS and reagent respectively, $\sigma_{ext}^{\neq}$ and $\sigma_{ext}$ are the external rotation symmetry numbers, for the TS and reagent respectively. $g_e^{\neq}$ and $g_e$ are the electronic state degeneracies of the transition state and reactant, respectively. $h$ is Planck's constant, $G^{\neq}(E - E_0)$ is the sum of states for the TS, $E_0$ is the reaction threshold energy and $\rho(E)$ is the reagent molecule state density.

The ME simulations were initiated from the chemically activated complex well, vdW1*. This species can evolve initially through three different pathways. (I) it can redissociate to reagents C + CH$_3$OCH$_3$ ; (II) it can surmount the barrier TS1 to form INT1; (III) it can stabilize to vdW1 through collisions with the carrier gas. Preliminary calculations showed that stabilization of the chemically activated complex was negligible over the range of pressures and temperatures used in the experiments, so pathway (III) is not expected to play a role in the current experiments. If pathway (II) is followed, INT1 is allowed to evolve further in the present simulations as the various other possible wells (INT2, INT3 and INT4), TSs (TS3, TS6, TS7, TS8, TS9, TS10) and product channels connected to vdW1 are also taken into consideration. The total rate constant $k(T, [M])$ is determined by expression (3)

$$k(T, [M]) = k_{capt}(T)\left(1 - f_{C+CH_3OCH_3}(T, [M])\right) \quad (3)$$

where $f_{C+CH_3OCH_3}(T, [M])$ is the fraction of vdW1* molecules that redissociate to reagent molecules at the end of the simulation, following pathway (I). In order to derive the branching fractions of the various individual product channels, the following expression was applied

$$f_P(T, [M])/\left(1 - f_{C+CH_3OCH_3}(T, [M])\right) \quad (4)$$

where $f_P(T, [M])$ is the fraction of product molecules in channel P formed at the end of the simulation.

The simulations were performed over a wide range of pressures (100 kPa - $1 \times 10^{-8}$ Pa) and over the experimental temperature range (50-296 K), to check for possible pressure dependent effects in the experiments. Here, a grain size of 1 cm$^{-1}$ was used to obtain better precision at low temperature and to be consistent with the results of the *ktools* calculations described above. Lennard-Jones parameters for the carrier gases Ar and N$_2$ were taken from the literature[51] while they were estimated for the well species INT1, INT2, INT3, and INT4 (($\sigma$ = 5.0 Å, $\varepsilon$ / k$_B$ = 300 K). The standard exponential-down model[52] was used to describe energy transfer.

Sums and densities of states for wells vdW1, INT1, INT2, INT3, and INT4 as well as transition states TS1, TS3, TS6, TS7, TS8, TS9 and TS10 were calculated by the *Densum* program.[47] As *Densum* is only suitable for obtaining these quantities from fixed TSs, the barrierless entrance channel and the dissociation channels INT1→P1, INT2→P8 and INT3→P6 were all treated with the *ktools* program (here, the energies and vibrational frequencies for the INT→P8 channel used in the *ktools* calculations were derived at the MRCI+Q/AVDZ level employing CASSCF wavefunctions as described in section 4.1), allowing us to obtain the sums and densities of states for these pathways. Reaction enthalpies ($\Delta H_{rxn}$(0 K)) were supplied by single point calculations at the DLPNO-CCSD(T)/AVTZ level, in addition to the zero-point corrected reaction critical energies (barrier heights with ZPE correction) where the ZPEs of the various species were derived from DFT calculations at the M06-2X/AVTZ level.

### 4.3 Rate constants

Several representative plots of the variation of the C($^3$P) VUV LIF signal as a function of time recorded at 127 K are displayed in Figure 2.

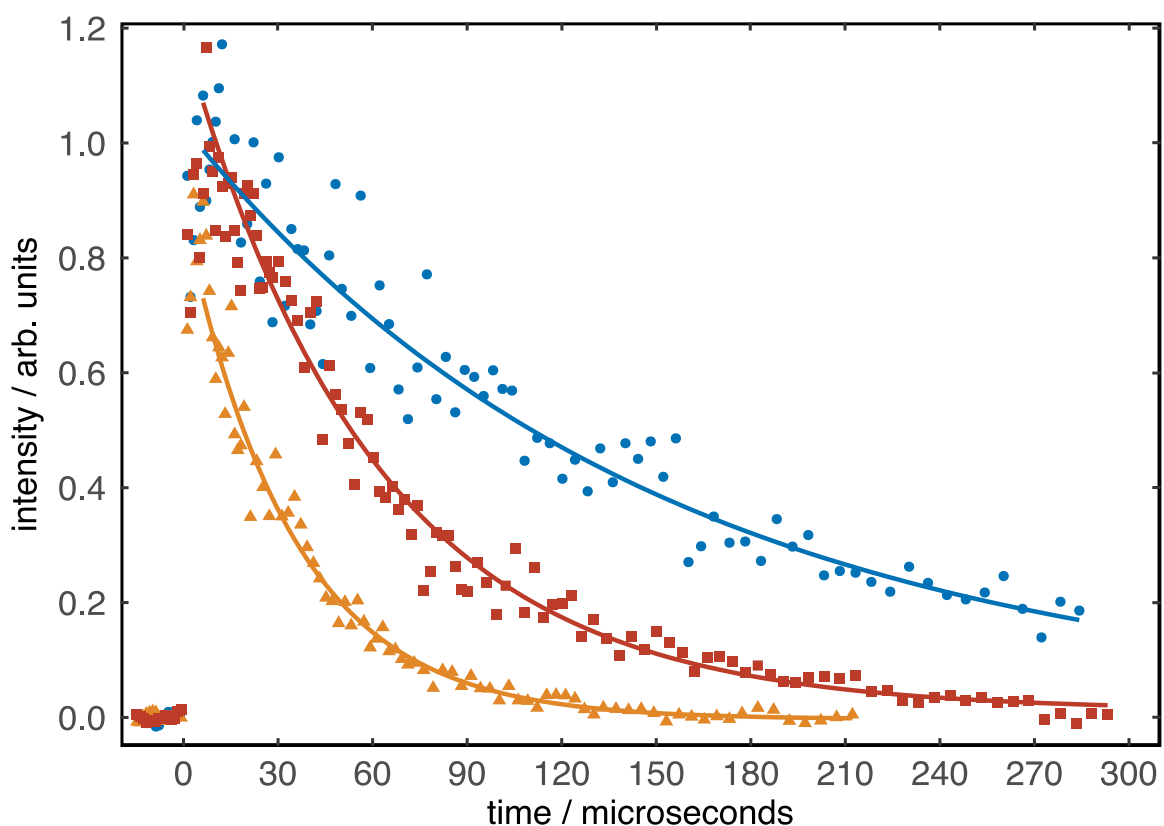

**Figure 2** C($^3$P) VUV LIF signal plotted as a function of delay time recorded at 127 K. (Blue circles) [CH$_3$OCH$_3$] = 0 cm$^{-3}$; (red squares) [CH$_3$OCH$_3$] = 2.0 × 10$^{14}$ cm$^{-3}$; (orange triangles) [CH$_3$OCH$_3$] = 6.0 × 10$^{14}$ cm$^{-3}$. The solid lines show the derived exponential fits to the individual datasets based on expression (5).

As CH$_3$OCH$_3$ concentrations were always much greater than the C($^3$P) atom concentration, the kinetic data could be analyzed assuming that the CH$_3$OCH$_3$ concentration did not change during the course of the reaction. In this respect, C($^3$P) atoms were considered to decay according to a first-order rate law (the so called pseudo-first-order approximation), obeying the following expression

$$I(t) = I_0 \exp(-k_{1st}t) \qquad (5)$$

$I(t)$ and $I_0$ are the time dependent and initial C($^3$P) VUV LIF intensities respectively (arbitrary units), $k_{1st}$ is the pseudo-first-order decay constant for C($^3$P) loss (s$^{-1}$) and $t$ is the delay time (s). The temporal dependence of the C($^3$P) fluorescence signal is well described by expression (5), decaying exponentially even in the absence of coreagent CH$_3$OCH$_3$. This occurs due additional reactive losses of C($^3$P) atoms such as through their reaction with CBr$_4$ molecules ($k_{C+CBr_4}[CBr_4]$) or impurities in the carrier gas ($k_{C+X}[X]$), in addition to their diffusive losses (diffusion out of the zone illuminated by the probe laser, $k_{diff}$). These three terms are constant for any single series of measurements, so the change in decay constant shown in figure 3 is due to the varying term $k_{C+CH_3OCH_3}[CH_3OCH_3]$ where $k_{C+CH_3OCH_3}$ is the second-order rate constant for the C($^3$P) + CH$_3$OCH$_3$ reaction. Consequently, $k_{C+CH_3OCH_3}$ can be determined at a given temperature by plotting the derived $k_{1st}$ values as a function of the corresponding CH$_3$OCH$_3$ concentration as shown in Figure 3.

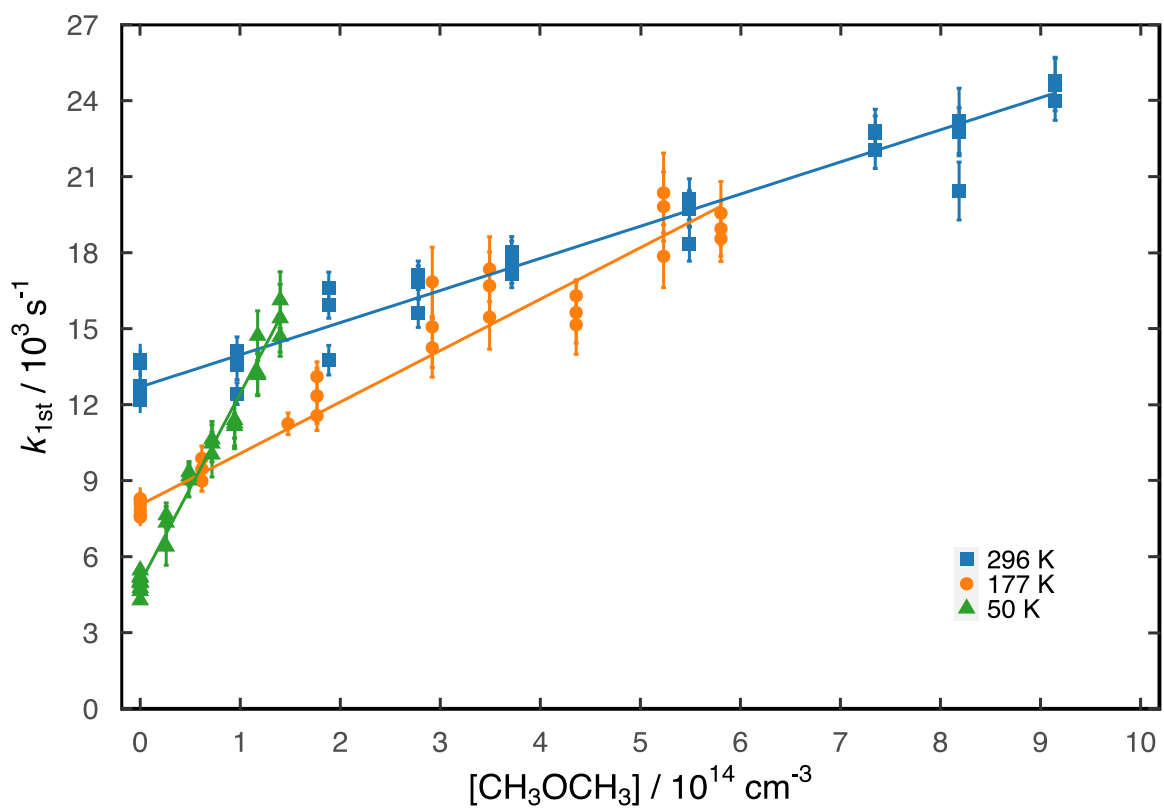

**Figure 3** Derived pseudo-first-order rate constants as a function of the CH$_3$OCH$_3$ concentration for experiments performed at various temperatures. (Blue solid squares) data recorded at 296 K; (orange solid circles) data recorded at 177 K; (green solid triangles) data recorded at 50 K. Solid lines represent weighted linear least-squares fits to the individual datasets yielding the second-order rate constants, $k_{\mathrm{C+CH_3OCH_3}}$, from the slopes.

The derived second-order rate constants are plotted as a function of temperature in Figure 4 and are summarized in Table 1 alongside other relevant information.

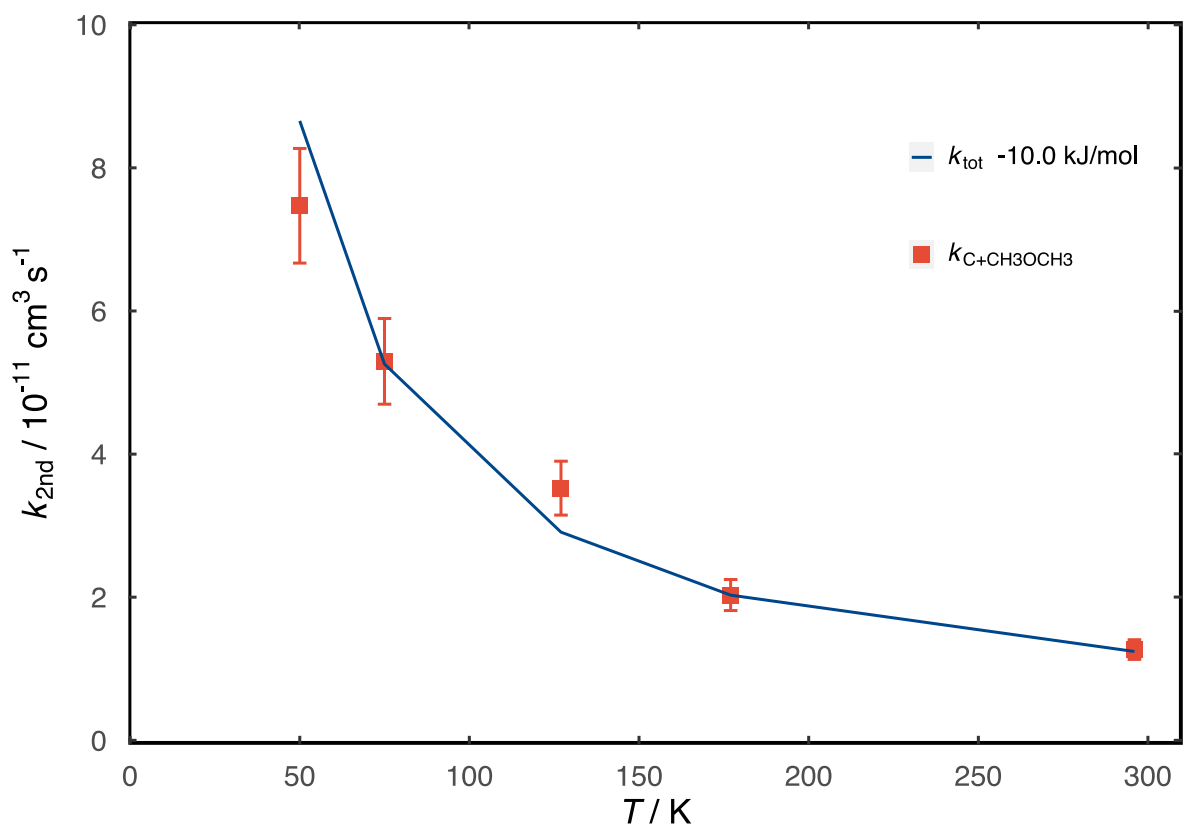

**Figure 4** Temperature dependence of the second-order rate constants for the C($^3$P) + CH$_3$OCH$_3$ reaction. (Red solid squares) this experimental work. Error bars represent the combined statistical (single standard deviation) and estimated systematic errors (10 %) of the nominal rate constant value. (Solid blue line) second-order rate constants calculated by the ME simulations assuming a submerged barrier of -10 kJ/mol for TS1.

**Table 1** Second-order rate constants for the C($^3$P) + CH$_3$OCH$_3$ reaction

| T / K | $N^b$ | Carrier gas | [CH$_3$OCH$_3$] / 10$^{13}$ cm$^{-3}$ | Flow density / 10$^{17}$ cm$^{-3}$ | $k_{C(^3P)+CH_3OCH_3}$ / 10$^{-11}$ cm$^3$ s$^{-1}$ | $k_{C(^3P)+CH_3OCH_3}$ / 10$^{-11}$ cm$^3$ s$^{-1}$ |
|---|---|---|---|---|---|---|
| 296 | 30 | Ar | 0 - 91.5 | 1.64 | (1.27 ± 0.14)$^c$ | 1.24$^d$ |
| 177 ± 2$^a$ | 28 | N$_2$ | 0 - 58.1 | 0.94 | (2.03 ± 0.22) | 2.03 |
| 127 ± 2 | 30 | Ar | 0 - 66.0 | 1.26 | (3.52 ± 0.38) | 2.91 |
| 75 ± 2 | 21 | Ar | 0 - 18.9 | 1.47 | (5.30 ± 0.60) | 5.26 |
| 50 ± 1 | 27 | Ar | 0 - 14.0 | 2.59 | (7.47 ± 0.80) | 8.66 |

[a]The uncertainties attributed to the calculated temperatures represent the statistical (1σ) errors obtained from the variations of the impact pressure as a function of distance obtained by Pitot tube measurements. [b]Number of individual measurements. [c]Measured rate constants. Uncertainties represent the combined statistical (1σ) and estimated systematic errors (10%). [d]Calculated rate constants obtained by adopting a value of -10.0 kJ/mol for TS1.

The C + $CH_3OCH_3$ reaction is seen to display a significant negative temperature dependence over the 50-296 K range with rate constants increasing by more than a factor of five from (1.3 ± 0.1) × $10^{-11}$ $cm^3$ $s^{-1}$ at 296 K to (7.5 ± 0.8) × $10^{-11}$ $cm^3$ $s^{-1}$ at 50 K. A simple fit to these data of the form k($T$) = α($T$/300)$^β$ can be used to represent the rate constants for this reaction that are appropriate for use in astrochemical models with α = (1.27 ± 0.40) × $10^{-11}$ and β = -1.01 ± 0.07. If we extrapolate the fit down to a typical dense interstellar cloud temperature of 10 K, a value of $k_{C+CH_3OCH_3}$(10 K) = 4.01 × $10^{-10}$ $cm^3$ $s^{-1}$ is obtained. While qualitatively similar temperature dependences were observed in earlier studies of the related C + $CH_3OH$[53] and C + $NH_3$ reactions,[36] the reactions of atomic carbon with acetonitrile, $CH_3CN$,[54] and acetone, $CH_3COCH_3$[55] display contrasting behaviour, being rapid over the entire temperature range between 50 and 296 K, with little or no temperature dependence. This difference appears to be due the nature of the TS separating the initial complex formed by the reagents from the more stable adducts over the respective potential energy surfaces. Indeed, while the C + $CH_3OH$ and C + $NH_3$ reactions are characterized by TSs that are close in energy to the reagent level, the C + $CH_3CN$ and C + $CH_3COCH_3$ reactions display TSs that are much lower in energy. As a result, once the initial complex is formed during the reactions of C with $CH_3OH$ and $NH_3$, it is equally likely to redissociate to reagents at higher temperature as it is to pass through to products. At low temperature, the forward process dominates due to the lower barrier for product formation. Although the C + $CH_3OCH_3$ reaction behaves in a similar manner to the C + $CH_3OH$ and C + $NH_3$ reactions, indicating the possible presence of a submerged TS, the results of the electronic structure calculations described in section 4.1 predict a real barrier to product formation in this case.

As a first attempt to simulate the experimental results through the kinetic calculations described in section 4.2, we used the nominal value of the TS1 activation barrier of + 14 kJ/mol obtained at the DLPNO-CCSD(T)/AVTZ//M06-2X/AVTZ level in test ME simulations performed at 296 K and at 50 K, in the absence of tunneling effects. Unsurprisingly, given the relatively high positive value of this TS, large values of $f_{\text{C}+\text{CH}_3\text{OCH}_3}(T,[M])$ were calculated at both temperatures, corresponding to more than 99.9 % of the initial vdW1* population redissociating to reagents C + CH₃OCH₃. When tunneling was included for these calculations, through the use of an unsymmetrical Eckart barrier, the results were the same as those performed in the absence of tunneling. At 296 K, the resulting total rate constant, $k(T,[M])$, was calculated to be several orders of magnitude lower than the measured value of (1.3 ± 0.1) × 10⁻¹¹ cm³ s⁻¹ at this temperature. As stabilization of vdW1 was already shown to be negligible at all temperatures for the pressures used in the present experiments, only a significant flux over TS1 can explain the magnitude of the measured rate constants. Consequently, the energy of TS1 was adjusted until good agreement with the experimental rate constants was obtained, while all other parameters including the vdW1 well depth were kept at their original values. The best agreement between theory and the experimental results over the 50-296 K range as shown in Figure 4 and summarized in Table 1 was obtained for a TS1 value of -10 kJ/mol, 24 kJ/mol lower than the value derived from the electronic structure calculations. In order to evaluate whether such a large change in the energy of TS1 is reasonable, we performed additional calculations of the ZPE corrected TS1 energy (relative to the C+ CH₃OCH₃ level) with a variety of methods. The results of these additional calculations are shown in Table 2.

**Table 2** Calculated TS1 energies corrected for ZPE effects relative to the C + CH₃OCH₃ reagent level.

| Method | TS1 energy / kJ mol⁻¹ |
| --- | --- |
| M06-2X/AVTZ | -8.8 |
| M06-2X/AVQZ[a] | -9.7 |
| M06-2X/CBS[b] | -6.3 |
| B3LYP/AVTZ | -21.0 |
| B2PLYP-D3/AVTZ | -12.2 |
| DLPNO-CCSD(T)/AVTZ//M06-2X/AVTZ[c] | +14.2 |
| DLPNO-CCSD(T)/CBS//M06-2X/CBS | +10.2 |
| CCSD(T)/AVTZ//M06-2X/AVTZ | +9.6 |
| DLPNO-CCSD(T)/AVTZ//B3LYP/AVTZ | +14.4 |
| CCSD(T)/AVTZ//B3LYP/AVTZ | +8.3 |
| DLPNO-CCSD(T)/AVTZ//B2PLYP-D3/AVTZ | +14.4 |

[a] aug-cc-pVQZ basis set. [b] Extrapolation to the complete basis set limit. [c]This method was used to obtain the energies displayed in Figure 1.

It can be seen that the DFT based methods in general lead to negative values of the TS1 energy including the M06-2X/CBS calculations where the results of calculations performed with M06-2X functional employing the AVTZ and AVQZ basis sets are extrapolated to the complete basis set limit. In contrast, the results of the coupled cluster based methods employing molecular geometries obtained at the DFT level all predict a positive value for the TS1 energy with the canonical CCSD(T) calculations leading to smaller barrier heights by a few kJ/mol than the more approximate DLPNO-CCSD(T) ones.

## 4.4 Product Branching Ratios

To obtain quantitative information regarding the product branching ratios of the $C(^3P)$ + $CH_3OCH_3$ reaction, separate experiments were performed to measure the product H-atom VUV LIF signal intensities as a function of reaction time. These relative intensities were then compared with those obtained from a reference process, the $C(^3P)$ + $C_2H_4$ reaction with a known H-atom yield of 0.92 ± 0.04 at 300 K,[56] allowing us to derive absolute H-atom product yields. Previous studies[57] have shown that the H-atom yield of the $C(^3P)$ + $C_2H_4$ reaction should not change significantly at the temperatures and pressures used in the present work. As $C(^1D)$ atoms are also generated during the photolysis of $CBr_4$ molecules at 266 nm, it was necessary to consider the possibility of H-atom production by the $C(^1D)$ + $C_2H_4$ and $C(^1D)$ + $CH_3OCH_3$ reactions. To lower the concentration of $C(^1D)$ atoms in the supersonic flow, $N_2$ was used as the carrier gas in experiments conducted at 296 K, while $N_2$ is already used as the carrier gas at 177 K. For the experiments performed at 75 K, where Ar is the carrier gas, a large concentration of $N_2$ ($1.5 \times 10^{16}$ cm$^{-3}$) was added to the flow. The results of earlier experiments to characterize the flow properties when adding large $N_2$ concentrations indicated a slight increase in the flow temperature of a few Kelvin.[58] The $C(^1D) + N_2 \rightarrow C(^3P) + N_2$ quenching reaction becomes more efficient as the temperature falls with measured rate constants of (5.3 ± 0.5) × 10$^{-12}$ cm$^3$ s$^{-1}$ at 296 K increasing to (10.7± 1.1) × 10$^{-12}$ cm$^3$ s$^{-1}$ at 75 K.[35] Consequently, any C($^1$D) atoms present in the flows are expected to be removed under our experimental conditions within the first few microseconds at all temperatures. Typical H-atom formation curves recorded for the C + CH$_3$OCH$_3$ (blue points) and C + C$_2$H$_4$ (red points) reactions are shown in Figure 5.

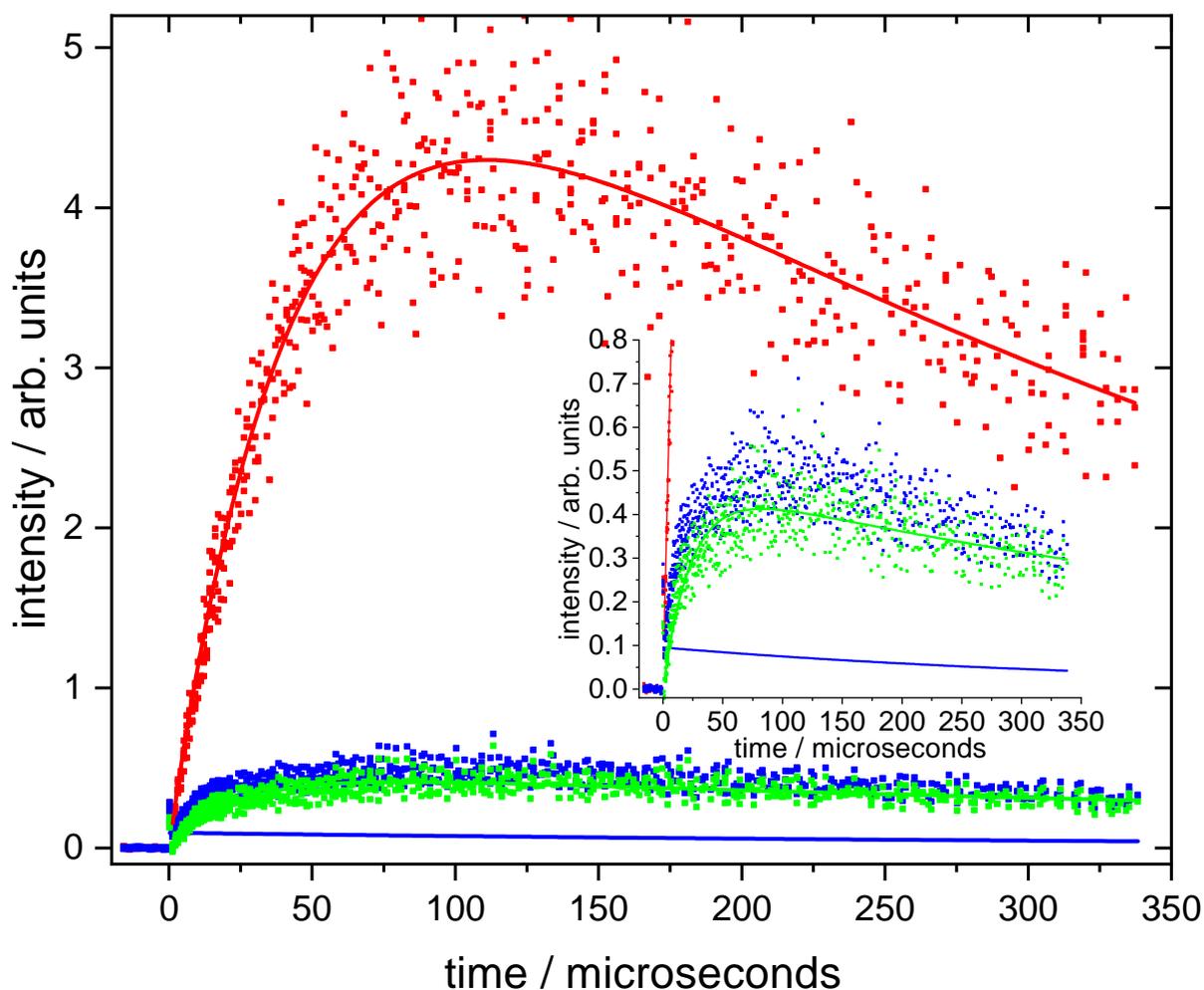

**Figure 5** H-atom VUV LIF signal intensity as a function of reaction time for the C + CH$_3$OCH$_3$ and C + C$_2$H$_4$ reactions recorded at 75 K. (Dark blue squares) H-atom signal from the reaction of both C($^3$P) and C($^1$D) with CH$_3$OCH$_3$ where [CH$_3$OCH$_3$] = 9.8 × 10$^{13}$ cm$^{-3}$. (Red solid squares) The C + C$_2$H$_4$ reaction with [C$_2$H$_4$] = 1.5 × 10$^{13}$ cm$^{-3}$. (Solid green squares) H-atom signal attributed to the C($^3$P) + CH$_3$OCH$_3$ reaction alone following subtraction of the instantaneous H-atom signal arising from the C($^1$D) + CH$_3$OCH$_3$ reaction. (Solid blue line) Estimated H-atom signal arising from the C($^1$D) + CH$_3$OCH$_3$ reaction. Solid red and green lines represent non-linear fits to the individual datasets using expression (6).

The curves plotted in Figure 5 are the result of several different experiments for both reactions, recorded alternately, which have then been coadded. While the H-atom formation curve for the C + $C_2H_4$ reaction shows a H-atom signal close to zero at short delay times, for the C + $CH_3OCH_3$ reaction it can be seen that there is an abrupt increase in the H-atom signal at short times followed by a slower rise at longer times. This rapid rise is likely to be due to the presence of unquenched C($^1$D) atoms present in the flow for the first few microseconds which then react with $CH_3OCH_3$ leading to a quasi-instantaneous production of atomic hydrogen which contributes to the overall H-atom signal. Similar contributions are also observed for those experiments performed at 177 and 296 K where $N_2$ was used as the carrier gas. This observation allows us to make certain hypotheses regarding the C($^1$D) + $CH_3OCH_3$ reaction. Firstly, it is likely that the rate constants for the C($^1$D) + $CH_3OCH_3$ reaction are considerably larger than those of the C($^3$P) + $CH_3OCH_3$ reaction, in a similar manner to the C($^3$P/$^1$D) + $CH_3OH$ reactions where the C($^1$D) + $CH_3OH$ reaction was measured to be ten times faster than the corresponding C($^3$P) + $CH_3OH$ reaction at 296 K with little or no temperature dependence.[53] Secondly, given the magnitude of the instantaneous signal, it could be that the H-atom yield of the C($^1$D) + $CH_3OCH_3$ reaction is much larger than the one we measure for the C($^3$P) + $CH_3OCH_3$ reaction. This conclusion was also reached during our earlier studies of the C($^1$D)+$CH_4$, $C_2H_6$ reactions[59] and during our study of the closely related C($^1$D) + $CH_3OH$ reaction[53] where H-atom yields were all close to unity. Despite the use of $N_2$ as the carrier gas, the instantaneous signal is larger at higher temperature due to the smaller rate constants of the C($^3$P) + $CH_3OCH_3$ reaction as the temperature increases. Indeed, the fact that the reaction rate is lower at higher temperature also means that larger $CH_3OCH_3$ concentrations are required to measure kinetic decays over similar timescales, while the difference in rate constants between the two processes is likely to be the greatest at higher temperature.

A similar issue is not expected to arise for the C($^3$P/$^1$D) + $C_2H_4$ reactions for several reasons. Firstly, the C($^3$P) + $C_2H_4$ is already very fast, with rate constants greater than $3 \times 10^{-10}$ cm$^3$ s$^{-1}$ at room temperature and below with only a weak temperature dependence,[9] meaning that lower concentrations of $C_2H_4$ are typically used for these experiments, reducing the contribution of the C($^1$D) + $C_2H_4$ reaction to the overall H-atom signal. Consequently, as the C($^1$D) + $C_2H_4$ reaction occurs on a much longer timescale, it is likely that most of the C($^1$D) atoms present in the flow following photolysis are already quenched to the ground state through collisions with the carrier gas before significant production of H-atoms occurs. Secondly, the H-atom yield of the C($^3$P) + C$_2$H$_4$ reaction is already close to unity, in contrast to the C($^3$P) + CH$_3$OCH$_3$ reaction, so that this signal almost certainly overwhelms the one from the C($^1$D) + C$_2$H$_4$ reaction.

As the H-atom signal contribution from the C($^1$D) + CH$_3$OCH$_3$ reaction occurs instantaneously, it can be subtracted from the overall signal relatively easily by estimating the H-atom signal at short delay times. As these H-atoms are expected to be lost mostly by diffusion from the observation region, this instantaneously produced H-atom signal is expected to decay exponentially as a function of time. The H + CH$_3$OCH$_3$ association reaction is not expected to play a role in H-atom loss due to the presence of an activation barrier for this process. An estimation of the decay time constant was made by examining the time constants of similar H-atom decay profiles obtained in earlier experiments under similar conditions. The final contribution to be subtracted from the H-atom production curves of the C + CH$_3$OCH$_3$ reaction for experiments performed at 75 K is shown as a solid blue line in Figure 5. The resulting subtraction shown as green squares is well described by the following biexponential function

$$I_H = A\{exp(-k_{L(H)}t) - exp(-k_{1st}t)\} \qquad (6)$$

where $I_H$ is H-atom fluorescence signal, $k_{L(H)}$ is the secondary H-atom loss term which occurs through processes such as diffusion with $k_{1st}$ equal to the first-order formation rate of atomic hydrogen. *A* is the predicted amplitude in the absence of H-atom losses (when $k_{L(H)}$ = 0, $exp(-k_{L(H)}t)$ = 1). The derived biexponential fit using expression (6) is shown as a solid green line in Figure 5. As small variations in the derived fitting parameters led to significant variations in predicted *A* value in the present analysis, we chose instead to compare the peak values given by the biexponential fits to the data. The peak intensities of the H-atom formation curves for the C + C$_2$H$_4$ reference reaction were divided by 0.92 to correct for the fact that the measured H-atom yields for this process are smaller than 1.[56] The peak intensities of the H-atom formation curves for the C + CH$_3$OCH$_3$ reaction were also corrected for VUV absorption losses in the present experiments as the high CH$_3$OCH$_3$ concentrations used here were estimated to result in absorption losses at 121.567 nm as high as 25 % in the worst case (experiments conducted at 296 K). The derived absolute H-atom yields for the C($^3$P) + CH$_3$OCH$_3$ reaction are listed in Table 3 alongside the product branching ratios predicted by the ME analysis described in section 4.2, using the same TS1 energy of -10 kJ/mol as used in the kinetic simulations. As mentioned earlier, the pathways originating from vdW2 are not considered due the very small predicted well depth which is well within the expected error bars of the calculations. No other energies were modified during the present calculations.

Hydrogen atoms and their accompanying molecular coproducts are clearly only minor products of the C + $CH_3OCH_3$ reaction considering the very low H-atom yield when compared with the H-atom yield of the C + $C_2H_4 \rightarrow C_3H_3$ + H reaction shown in red in Figure 5.

**Table 3** Measured and calculated branching ratios (%) of the various reaction channels for the C($^3$P) + $CH_3OCH_3$ reaction. (The experimental measurement (Expt.) represents the sum of all H-atom production channels, which should be compared to the sum of P4 and P8.)

| T / K | P2 | P1 | P6 | P8 | P4 | P8+P4 | Expt. |
|---|---|---|---|---|---|---|---|
| 50 | 67.4 | 28.8 | 0.1 | 1.7 | 2.1 | 3.8 | |
| 75 ± 2[a] | 67.3 | 28.9 | 0.1 | 1.7 | 2.1 | 3.8 | 8.9 |
| 127 | 67.6 | 28.6 | 0.1 | 1.7 | 2.0 | 3.7 | |
| 177 ± 2 | 67.9 | 28.3 | 0.1 | 1.7 | 2.0 | 3.7 | 7.8±5.3[b] |
| 296 | 68.5 | 27.7 | 0.1 | 1.7 | 2.0 | 3.7 | 9.9±4.9 |

[a]The uncertainties attributed to the calculated temperatures represent the statistical (1σ) errors obtained from the variations of the impact pressure as a function of distance obtained by Pitot tube measurements. [b]Errors represent the 95% confidence level (no error is cited for the measurement performed at 75 K as this is the result of a single series of coadded measurements).

Table 3 lists only those product channels with a non-zero yield predicted by the ME calculations (the yield for products P7, H + $CH_3COCH_2$, for example was always zero in the present simulations, while the yields for products P3 and P5 were also zero because the pathways from vdW2 were not considered). The principal result of these simulations is that pathways leading to products of lower molecular complexity dominate for the C + $CH_3OCH_3$ reaction. It can be seen that pathway INT1→TS3→P2 leading to the formation of $CH_3$ + $CH_3$ + CO is the major channel, representing approximately 67-68 % of the total according to our calculations, with little predicted variation as a function of temperature. The pathway INT1→P1 leading to the formation of $CH_3OC$ + $CH_3$ is also important (28-29%) although it is probable that a significant fraction of the $CH_3OC$ formed here will dissociate further to $CH_3$ +

CO. The INT1→TS9→INT2→TS6→INT3→P6 channel leading to the formation of $CH_3$ + $CH_3CO$ is predicted to be essentially negligible, representing only 0.1 % of the calculated overall product yield. Although the major outcome of this reaction appears to be the formation of less complex products, there are two other minor channels INT1→TS9→INT2→P4 (H + $CH_3OCCH_2$) and INT1→TS9→INT2→P8 (H + $CH_2OCCH_3$) predicted by the ME calculations that lead to an increase in molecular complexity and the elimination of an H-atom. As the experiments measure the sum of all H-atom production channels, the measured H-atom yields should be compared to the sum of the calculated H-atom yields P4 and P8 as shown in Table 3. In order to correctly account for these C-H bond dissociation channels, where three C-H bonds can dissociate for each of the two $CH_3$ groups, we included three equivalent bond dissociation pathways in the ME calculations for both INT2→P4 and INT2→P8. Additional constrained optimizations from the $CH_3OCCH_3$ intermediate at the M06-2X/AVTZ level, performed by increasing the C-H distance from its equilibrium value to the separated products P4 for an alternative C-H bond showed that these pathways are effectively equivalent. These channels are calculated to represent approximately 4 % (this falls to 3 % if only one C-H dissociation channel is included for each $CH_3$ group) of the overall product yield of this reaction, in reasonable agreement with the low measured H-atom yields. In this respect, the experiments and calculations show good qualitative agreement by predicting that the channels leading to an increase in molecular complexity (those accompanied by the production of H-atoms) are clearly only a minor product of the C + $CH_3OCH_3$ reaction, while this reaction could be a source of $CH_3$ and CO in the interstellar medium.

## 5 Astrophysical Implications

Current astrochemical models largely underestimate the abundance of $CH_3OCH_3$ in dense clouds,[60] even before the introduction of the C + $CH_3OCH_3$ reaction. Here, we test the effects induced by including this reaction in a standard astrochemical model, the Nautilus code[61] (a 3-phase gas, dust grain ice surface and dust grain ice mantle time dependent chemical model employing kida.uva.2014[29] as the basic reaction network) updated recently for a better description of COM chemistry on interstellar dust grains and in the gas-phase.[62, 63] There are 800 individual species included in the network that are involved in approximately 9000 separate reactions. Elements are either initially in their atomic or ionic forms in this model (elements with an ionization potential < 13.6 eV are considered to be fully ionized) and the C/O elemental ratio is equal to 0.71 in this work. The initial simulation parameters are listed in Table 4.

**Table 4** Elemental abundances and other model parameters

| Element | Abundance[a] | $n_H + 2n_{H_2}$ / cm$^{-3}$ | T/ K | Cosmic ray ionization rate / s$^{-1}$ | Visual extinction |
|---|---|---|---|---|---|
| $H_2$ | 0.5 | $2.5 \times 10^4$ | 10 | $1.3 \times 10^{-17}$ | 10 |
| He | 0.09 | | | | |
| $C^+$ | $1.7 \times 10^{-4}$ | | | | |
| N | $6.2 \times 10^{-5}$ | | | | |
| O | $2.4 \times 10^{-4}$ | | | | |
| $S^+$ | $6.0 \times 10^{-7}$ | | | | |
| $Fe^+$ | $2.0 \times 10^{-8}$ | | | | |
| $Cl^+$ | $1.0 \times 10^{-7}$ | | | | |
| F | $6.7 \times 10^{-9}$ | | | | |

[a]Relative to total hydrogen ($n_H + 2n_{H_2}$)

The grain surface and the mantle are both chemically active for these simulations, while accretion and desorption are only allowed between the surface and the gas-phase. The dust-to-gas ratio (in terms of mass) is 0.01. A sticking probability of 1 is assumed for all neutral species while desorption occurs by both thermal and non-thermal processes (cosmic rays, chemical desorption) including sputtering of ices by cosmic-ray collisions.[64] The formalism used to describe the surface reactions and a more detailed description of the simulations can be found in Ruaud et al.[61]

The reactions involved in $CH_3OCH_3$ production and destruction are shown in Table S5 of the supplementary information file. In our updated network, four reactions lead to $CH_3OCH_3$ formation. First, there is the radiative association reaction between $CH_3$ and $CH_3O$.[60] Second, the electronic dissociative recombination of $CH_3OHCH_3^+$ also leads to $CH_3OCH_3$ production.[65-67] Two other reactions take place on the grains, firstly the s-$CH_3$ + s- $CH_3O$ reaction (s- means species on the grain) and finally the C + s- $CH_3OH$ reaction followed by hydrogenation of s-CH$_3$OCH. To estimate the efficiency of CH$_3$OCH$_3$ production, we turn off the reactions one by one. It appears that the gas-phase reactions (CH$_3$ + CH$_3$O and CH$_3^+$ + CH$_3$OH) are inefficient and contribute almost negligibly to CH$_3$OCH$_3$ production. The s-CH$_3$ + s-CH$_3$O route is not efficient at low temperatures, since at 10 K neither s-CH$_3$ nor s-CH$_3$O are predicted to diffuse on ice using an E$_{diffusion}$ = (0.3-0.4)×E$_{binding}$.[68] Most of the CH$_3$OCH$_3$ present in the gas-phase arises from cosmic ray-induced desorption of s-CH$_3$OCH$_3$ produced by the C + s-CH$_3$OH (+ 2 s-H) reaction.

The main losses of gas-phase CH$_3$OCH$_3$ are through its reactions with He$^+$, H$_3^+$ and HCO$^+$, through reactions with atomic C (this work) and the OH radical, and by depletion onto grains. The current model (dashed red line) underestimates the observed (hatched red rectangle) gas-phase abundance of CH$_3$OCH$_3$ shown in Figure 6. It should be noted that Nautilus uses the Minissale et al[69] formalism for chemical desorption, which is based on experiments. As such, in the present simulations, only a very low percentage of CH$_3$OCH$_3$ molecules produced by the s-H + s-CH$_3$OCH$_2$ -> CH$_3$OCH$_3$ reaction actually leads to desorption (10$^{-4}$ %). The chemical desorption level would need to be approximately 3 % for the simulated CH$_3$OCH$_3$ abundance to reproduce the observed one. However, there are other non-thermal processes that could also be more efficient as well. As an example, for cosmic-ray induced sputtering, we use the experimental yields for pure water ices while measured yields for pure CO$_2$ ices are higher (see discussion in Wakelam et al. 2021).[64] This suggest that the yield for mixed ices is probably higher than the one we use.

The solid red line in Figure 6 shows the nominal abundance of CH$_3$OCH$_3$ generated by the model in the absence of the C + CH$_3$OCH$_3$ reaction.

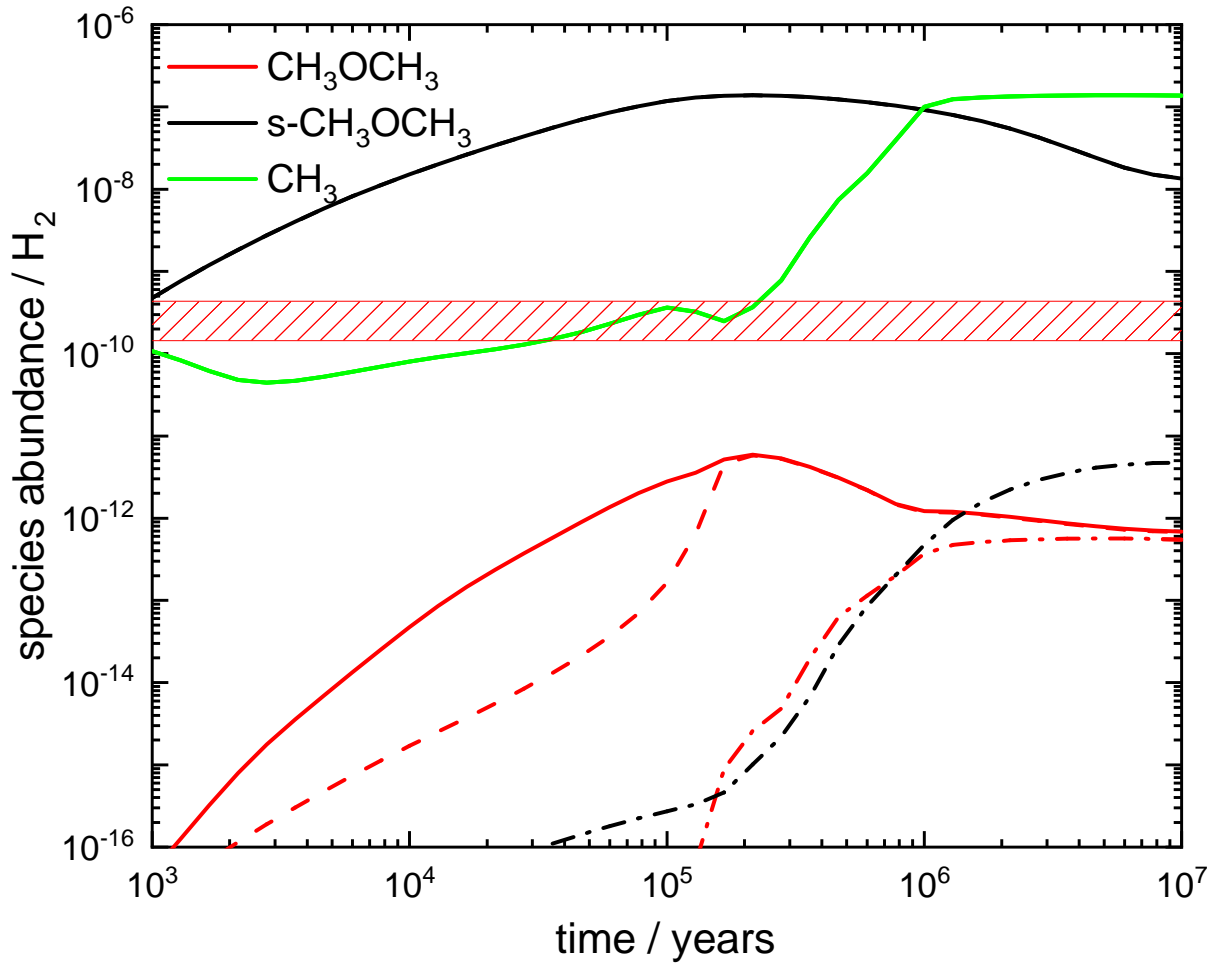

**Figure 6** Gas-grain astrochemical model results for the formation of gas-phase CH$_3$OCH$_3$ (red lines), solid phase CH$_3$OCH$_3$ (black lines) and gas-phase CH$_3$ (green lines) in dark clouds as a function of cloud age. (Solid lines) standard network results. (dashed lines) the same network with the gas-phase C + CH$_3$OCH$_3$ reaction included. (Dashed-dotted line) new network without the C + s-CH$_3$OH reaction. The horizontal red hatched rectangle represents the observed CH$_3$OCH$_3$ abundance in TMC-1[28] with an arbitrary error associated ($\pm\sqrt{3}$).

The inclusion of the C + CH$_3$OCH$_3$ reaction in the network with an estimated rate constant, $k_{\text{C}+\text{CH}_3\text{OCH}_3}(10\text{K}) = 1.3 \times 10^{-10}$ cm$^3$ s$^{-1}$ leads to a maximum decrease of the CH$_3$OCH$_3$ abundance of more than an order of magnitude between 10$^3$ and 10$^5$ years as shown by the dashed red line in Figure 6. Increasing the estimated rate constant to $4.0 \times 10^{-10}$ cm$^3$ s$^{-1}$, the value obtained by extrapolating the experimental fit to 10 K, leads to an additional threefold decrease of the CH$_3$OCH$_3$ abundance at short times. Given the experimental and theoretical results presented above, the formation of CH$_3$ + CH$_3$ + CO was assumed to be the only product channel of this reaction during these simulations. Despite this, and the fact that two CH$_3$ radicals are produced for each $CH_3OCH_3$ consumed, the additional $CH_3$ produced by this reaction is small compared to other sources of $CH_3$ in dense interstellar clouds as the inclusion of this reaction induces a negligible change in the simulated $CH_3$ abundance (the solid and dashed green lines are indistinguishable in Figure 6). This reaction is also a negligible source of CO compared to other pathways already present in the model. At ages considered to be characteristic of typical dense clouds such as TMC-1 (a few $10^5$ years), atomic carbon is removed from the gas-phase through reactions forming CO and by accretion onto grains, thereby limiting the effect of the C + $CH_3OCH_3$ reaction at longer times (>$10^5$ years). Despite the large effect of the C + $CH_3OCH_3$ reaction on $CH_3OCH_3$ abundances at early times, these simulations indicate that the C + $CH_3OCH_3$ reaction induces only small changes in the gas-phase $CH_3OCH_3$ abundance at typical dense interstellar cloud ages, with a calculated $CH_3OCH_3$ abundance that underestimates the observed one for TMC-1. When the C + s-$CH_3OH$ reaction is switched off, both the gas-phase (red dashed-dotted line) and solid phase (black dashed-dotted line) $CH_3OCH_3$ abundances fall by several orders of magnitude, showing the crucial role of this reaction. This also demonstrates the importance of Eley-Rideal reactions for $CH_3OCH_3$, which, along with $CH_3OH$, $C_2H_5OH$, and $H_2S$, is a compound produced almost exclusively on grains. Its detection therefore enables us to better characterize the chemistry on grains in dense, cold regions, where COMs cannot a priori be synthesized in the same way as in protostars, since only atomic and molecular hydrogen are expected to be mobile on ice at 10 K. Furthermore, chemical desorption is assumed to be relatively inefficient for $CH_3OCH_3$ due to its size[69] and high binding energy (4080 ± 500 K[70]), so its abundance in the gas phase is also a means of assessing the efficiency of desorption induced by cosmic-ray collisions.[64]

## 6 Conclusions

This work describes the results of an experimental investigation of the C($^3$P) + $CH_3OCH_3$ reaction, supported by theoretical calculations of the rate constant and product channels, followed by an astrochemical modeling study of its effects on interstellar chemistry. Experimentally, a supersonic flow reactor was used to investigate this process over the 50-296 K range. Pulsed laser photolysis was used for the in-situ generation of C($^3$P) atoms, which were followed by pulsed laser induced fluorescence in the vacuum ultraviolet wavelength range. Theoretically, electronic structure calculations were performed on the ground $^3$A′ surface of $C_3H_6O$ to determine the various intermediates, transition states and product channels involved. Based on the derived energies and structures, a master equation analysis was used to calculate rate constants and product branching ratios for this reaction as a function of temperature. The reaction is seen to accelerate as the temperature falls, while the low measured H-atom yields at all temperatures support the theoretical prediction that the major products should be $CH_3$ + $CH_3$ + CO. The C($^3$P) + $CH_3OCH_3$ reaction was introduced into a gas-grain model of dense interstellar clouds. The simulations predict that $CH_3OCH_3$ abundances decrease significantly at intermediate times when gas-phase C($^3$P) abundances are high, while the predicted $CH_3OCH_3$ abundances only reach high levels at later cloud ages when most of the C($^3$P) has been locked up in interstellar reservoirs such as CO. Despite its high reactivity, the C($^3$P) + $CH_3OCH_3$ reaction is likely to be only a very minor source of $CH_3$ in dense interstellar clouds.

**Author information**

Corresponding Author

Kevin M. Hickson – Institut des Sciences Moléculaires ISM, CNRS UMR 5255, Univ. Bordeaux, F-33400 Talence, France; orcid.org/0000-0001-8317-2606;

Email: kevin.hickson@u-bordeaux.fr

Authors

Jean-Christophe Loison – Institut des Sciences Moléculaires ISM, CNRS UMR 5255, Univ. Bordeaux, F-33400 Talence, France

Valentine Wakelam – Laboratoire d'astrophysique de Bordeaux, CNRS, Univ. Bordeaux, F 33615 Pessac, France

**Supporting Information**

Geometries and frequencies of the stationary points involved in the C($^3$P) + $CH_3OCH_3$ reaction obtained at the M06-2X/aug-cc-pVTZ level of theory (DOCX). Energies, frequencies and rotational constants along the entrance channel leading to vdW1 formation. Reactions involved in $CH_3OCH_3$ production and destruction in the astrochemical network.

**Acknowledgements**

K.M.H. acknowledges support from the French program 'Physique et Chimie du Milieu Interstellaire'' (PCMI) of the CNRS/INSU with the INC/INP cofunded by the CEA and CNES as well as funding from the 'Program National de Planétologie'' (PNP) of the CNRS/INSU.

Supporting information file for

**A Low Temperature Kinetic Study of the C($^3$P) + CH$_3$OCH$_3$ Reaction. Rate constants, H-atom Product Yields and Astrochemical Implications**

Kevin M. Hickson,[1,*] Jean-Christophe Loison,[1] and Valentine Wakelam[2]

[1]Institut des Sciences Moléculaires ISM, CNRS UMR 5255, Univ. Bordeaux, 351 Cours de la Libération, F-33400, Talence, France
[2]Laboratoire d'astrophysique de Bordeaux, CNRS, Univ. Bordeaux, B18N, allée Geoffroy Saint-Hilaire, F-33615 Pessac, France

**Geometries and frequencies of the stationary points involved in the C($^3$P) + CH$_3$OCH$_3$ reaction obtained at the M06-2X/aug-cc-pVTZ level of theory.**

**Geometries in Cartesian coordinates**

**Reactants and products**

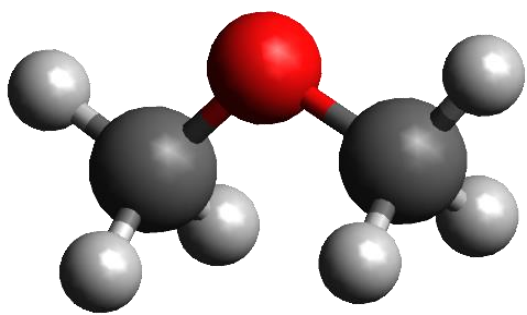

CH$_3$OCH$_3$

| | | | |
|---|---|---|---|
| O | 0.000001 | 0 | 0.593687 |
| C | 0 | 1.161471 | -0.19451 |
| C | 0 | -1.16147 | -0.19451 |
| H | 0 | 2.018972 | 0.47384 |
| H | -1E-06 | -2.01897 | 0.47384 |
| H | 0.889393 | 1.203788 | -0.83341 |
| H | -0.8894 | 1.203787 | -0.83341 |
| H | -0.88939 | -1.20379 | -0.83341 |
| H | 0.889395 | -1.20379 | -0.83341 |

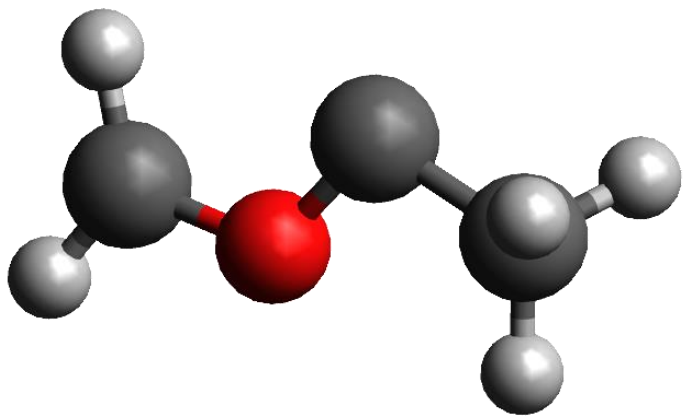

CH$_2$OCCH$_3$

| | | | |
|---|---:|---:|---:|
| O | 0.551426 | -0.39739 | -0.27947 |
| C | 2.746494 | 0.172572 | -0.24953 |
| H | 3.135243 | 1.051678 | 0.269798 |
| H | 2.534392 | 0.406786 | -1.29474 |
| H | 3.541063 | -0.5747 | -0.17941 |
| C | -0.62904 | -0.84545 | 0.219614 |
| H | -1.44037 | -0.86737 | -0.48349 |
| H | -0.65566 | -1.12591 | 1.258285 |
| C | 1.58632 | -0.33807 | 0.533893 |

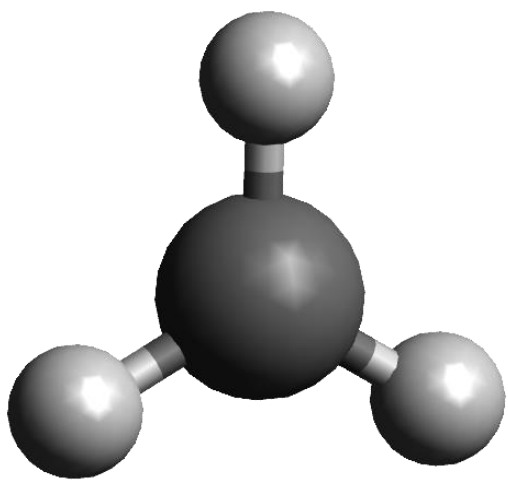

CH$_3$

| | | | |
|---|---:|---:|---:|
| C | -0.70372 | 1.186541 | -5.8E-05 |
| H | -0.70393 | 0.222151 | -0.4783 |
| H | -0.70309 | 2.083072 | -0.5959 |
| H | -0.70309 | 1.254814 | 1.074252 |

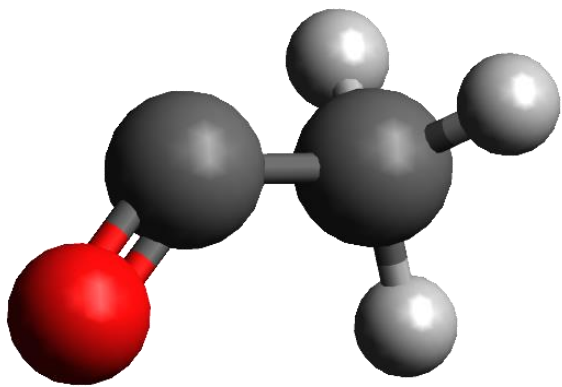

CH₃CO

| | | | |
|---|---|---|---|
| O | -0.74553 | 1.29281 | -0.37691 |
| C | 1.59815 | 0.710808 | -0.23507 |
| H | 2.270581 | 1.396212 | 0.277147 |
| H | 1.66455 | 0.838353 | -1.31566 |
| H | 1.875973 | -0.30075 | 0.053802 |
| C | 0.194665 | 0.976079 | 0.248945 |

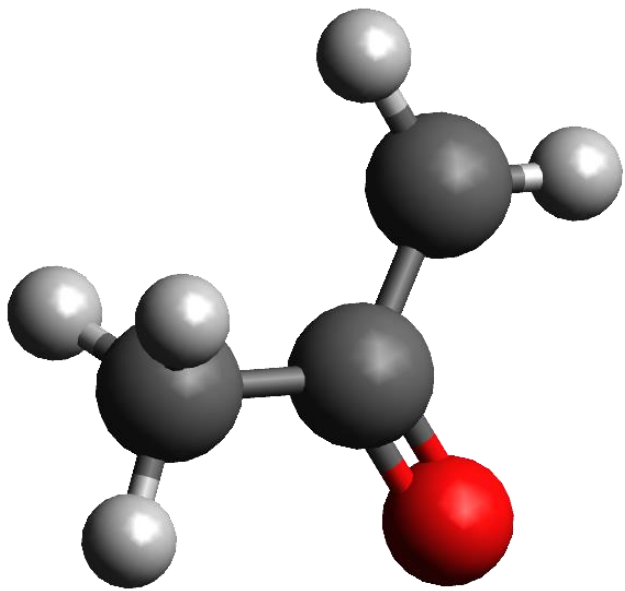

CH₃COCH₂

| | | | |
|---|---|---|---|
| C | -0.00086 | 0.148935 | -1E-06 |
| O | 0.427942 | 1.295073 | 0 |
| C | -1.41953 | -0.09936 | 0 |
| C | 0.925689 | -1.04507 | 0 |
| H | -2.09164 | 0.744823 | 0.000001 |
| H | -1.81879 | -1.10279 | 0.000001 |
| H | 1.956691 | -0.70529 | 0.000001 |
| H | 0.742514 | -1.66434 | 0.878793 |
| H | 0.742516 | -1.66434 | -0.87879 |

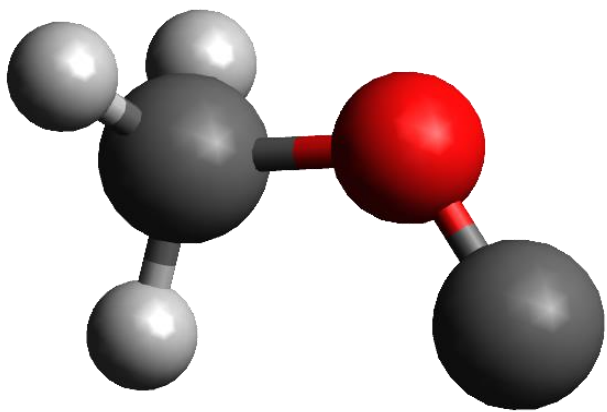

CH₃OC
| | | | |
|---|---|---|---|
| O | -1.19467 | 0.827273 | 0.165282 |
| C | -1.2939 | -0.63672 | 0.061527 |
| H | -2.33199 | -0.89373 | 0.245446 |
| H | -0.97396 | -0.9334 | -0.93311 |
| H | -0.65161 | -1.0475 | 0.833542 |
| C | -0.82459 | 1.5167 | -0.81057 |

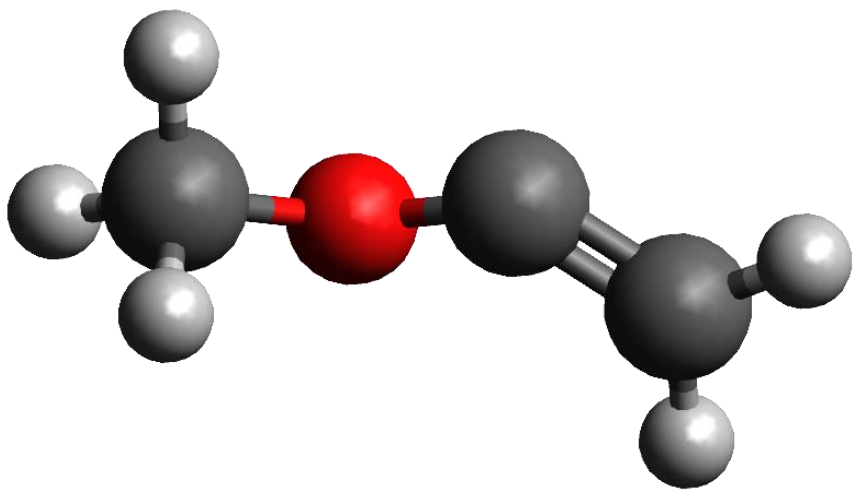

CH₃OCCH₂
| | | | |
|---|---|---|---|
| O | 0.161042 | -0.23304 | -0.30566 |
| C | 2.429567 | 0.532216 | -0.17579 |
| H | 3.184404 | 0.937174 | 0.479557 |
| H | 2.650462 | 0.432194 | -1.23393 |
| C | -0.47068 | -1.37802 | 0.268988 |
| H | -1.3652 | -1.56093 | -0.31741 |
| H | 0.19694 | -2.23791 | 0.217179 |
| H | -0.73476 | -1.17677 | 1.306307 |
| C | 1.263118 | 0.162634 | 0.300079 |

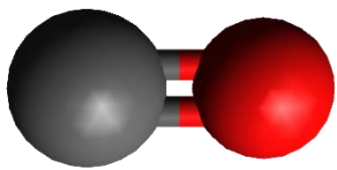

CO

| C | 0 | 0 | -0.64096 |
| O | 0 | 0 | 0.48043 |

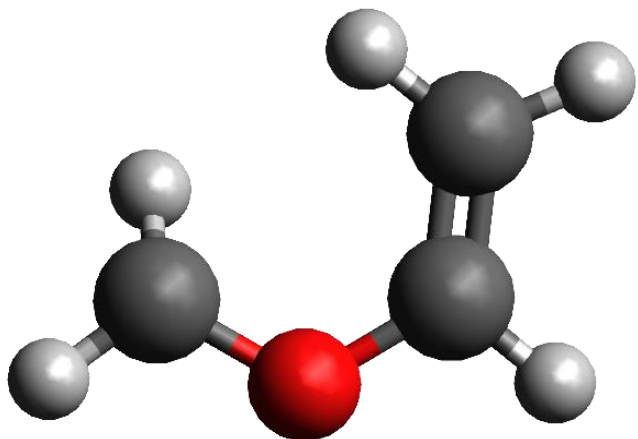

H₂CCHOCH₂

| O | -0.05854 | 0.727316 | -0.92497 |
| C | 1.063769 | 1.123118 | -0.28273 |
| H | 1.518344 | 0.429444 | 0.410454 |
| H | -0.22661 | 0.024127 | 1.710468 |
| H | 1.648896 | 1.827412 | -0.84876 |
| C | -0.92978 | -0.0959 | -0.28051 |
| H | -1.66007 | -0.48288 | -0.97628 |
| C | -0.93135 | -0.39617 | 1.009659 |
| H | -1.6862 | -1.07023 | 1.381944 |

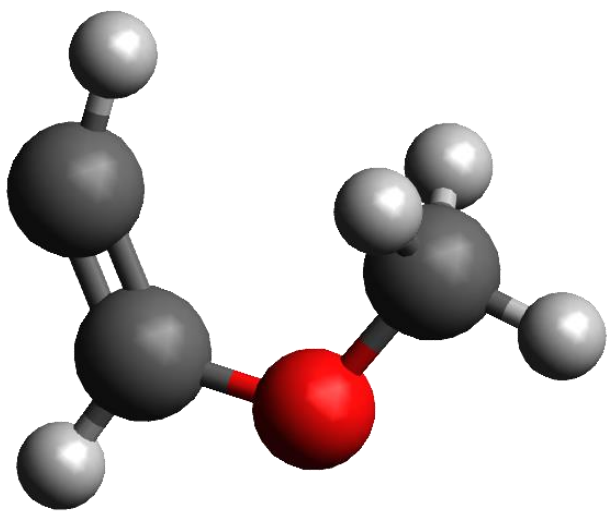

HCCHOCH₃

| | | | |
|---|---|---|---|
| O | -0.24377 | 0.83835 | -0.81111 |
| C | 0.792658 | 0.960075 | 0.140836 |
| H | 1.322449 | 0.011188 | 0.255287 |
| H | 0.390034 | 1.263507 | 1.110362 |
| H | 1.472095 | 1.720253 | -0.2315 |
| C | -1.19959 | -0.07796 | -0.50774 |
| H | -1.95733 | -0.1116 | -1.28173 |
| C | -1.26748 | -0.85071 | 0.54927 |
| H | -0.73935 | -1.07846 | 1.455183 |

**Intermediates**

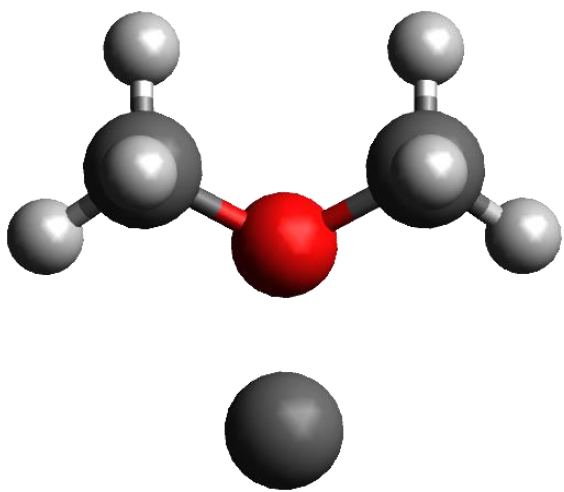

vdW1

| | | | |
|---|---|---|---|
| O | -0.26659 | 0.54844 | 0.21022 |
| C | 1.14860 | 0.36722 | 0.10533 |
| H | 1.59007 | 1.35258 | 0.20868 |
| H | 1.38446 | -0.05085 | -0.87365 |
| H | 1.47375 | -0.29362 | 0.90761 |
| C | -1.00595 | -0.67221 | 0.10351 |
| H | -2.05170 | -0.40467 | 0.20867 |
| H | -0.82900 | -1.11479 | -0.87729 |
| H | -0.69111 | -1.34134 | 0.90309 |
| C | -0.79345 | 1.63044 | -0.91095 |

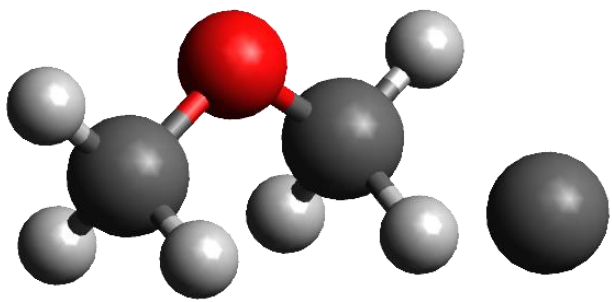

vdW2

| | | | |
|---|---|---|---|
| O | -0.51334 | -0.49696 | -0.44761 |
| C | 0.57429 | -0.73610 | 0.35195 |
| H | 1.27161 | -1.39343 | -0.15486 |
| H | 0.30726 | -1.11509 | 1.34673 |
| H | 1.12119 | 0.24206 | 0.63062 |
| C | -1.42541 | 0.41718 | 0.12415 |
| H | -2.24191 | 0.53999 | -0.58097 |
| H | -1.81405 | 0.03554 | 1.07330 |
| H | -0.93822 | 1.38224 | 0.29170 |
| C | 1.93374 | 1.16608 | -0.22665 |

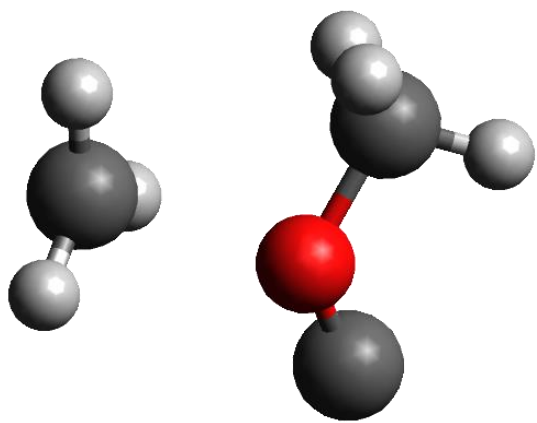

INT1

| | | | |
|---|---|---|---|
| O | -0.24338 | 0.032625 | -0.41511 |
| C | 3.016732 | -0.40236 | -0.16901 |
| H | 2.80868 | 0.226502 | 0.680798 |
| H | 2.84071 | -0.0254 | -1.16271 |
| H | 3.522668 | -1.34371 | -0.03366 |
| C | -0.14455 | -1.29644 | 0.207091 |
| H | -1.14962 | -1.70537 | 0.229 |
| H | 0.507696 | -1.88828 | -0.42708 |
| H | 0.26989 | -1.17913 | 1.204689 |
| C | 0.0974 | 1.060938 | 0.209318 |

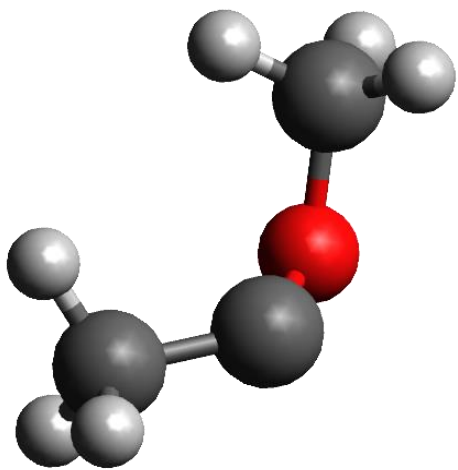

INT2

| | | | |
|---|---|---|---|
| O | 0.175388 | -0.18571 | -0.29806 |
| C | 2.659922 | 0.025238 | -0.18774 |
| H | 3.37031 | 0.546367 | 0.449877 |
| H | 2.812139 | 0.362351 | -1.21857 |
| H | 2.891744 | -1.04549 | -0.14677 |
| C | -0.31777 | -1.40382 | 0.265628 |
| H | -1.19506 | -1.6856 | -0.3096 |
| H | 0.439882 | -2.18577 | 0.194679 |
| H | -0.58735 | -1.24886 | 1.309958 |
| C | 1.277035 | 0.300663 | 0.263914 |

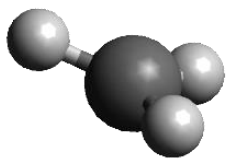

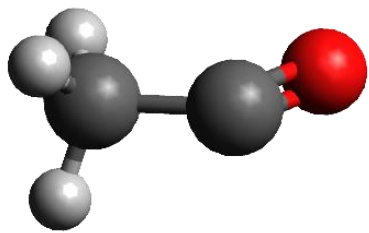

INT3

| | | | |
|---|---|---|---|
| O | 0.517289 | 1.374719 | -0.0459 |
| C | 2.46384 | -0.06113 | -0.1464 |
| H | 3.404359 | 0.487115 | -0.09612 |
| H | 2.20977 | -0.29273 | -1.18109 |
| H | 2.568706 | -0.97075 | 0.439229 |
| C | -0.23121 | -1.9529 | 0.214698 |
| H | -0.69881 | -1.30552 | -0.50756 |
| H | 0.183244 | -2.89778 | -0.09438 |
| H | -0.29303 | -1.71062 | 1.262621 |
| C | 1.402083 | 0.808978 | 0.478215 |

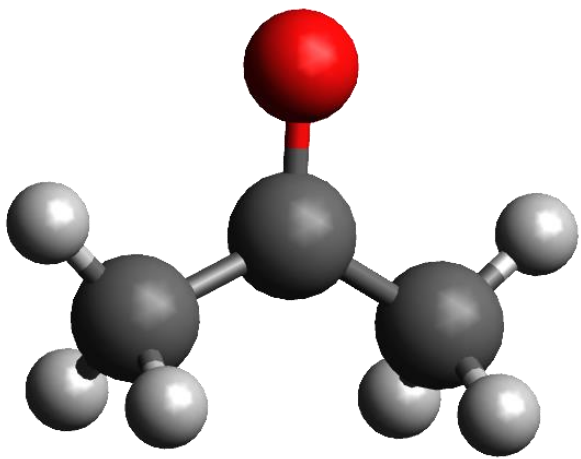

INT4

| | | | |
|---|---|---|---|
| O | -0.52382 | 0.000035 | 1.286584 |
| C | 0.039028 | 1.297329 | -0.6047 |
| H | 0.2591 | 2.14577 | 0.039953 |
| H | -0.97073 | 1.425227 | -1.0119 |
| H | 0.750732 | 1.28674 | -1.42789 |
| C | 0.038923 | -1.29734 | -0.60467 |
| H | 0.25892 | -2.14578 | 0.039992 |
| H | -0.97085 | -1.42516 | -1.01188 |
| H | 0.750631 | -1.28682 | -1.42786 |
| C | 0.157373 | -1E-06 | 0.152863 |

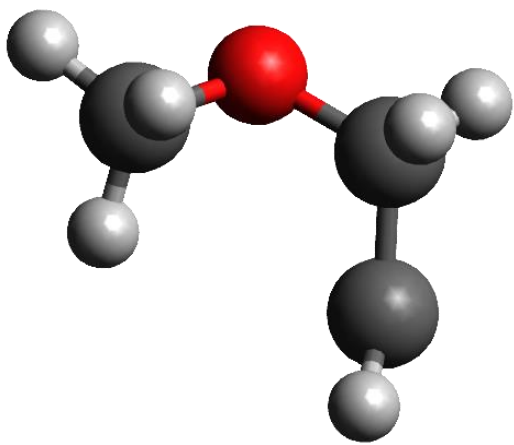

INT5

| | | | |
|---|---|---|---|
| O | -0.05182 | 0.63939 | -0.82779 |
| C | 0.944503 | 0.932093 | 0.121977 |
| H | 1.694207 | 0.133354 | 0.164145 |
| H | 0.520113 | 1.069367 | 1.120795 |
| H | 1.425671 | 1.855491 | -0.18964 |
| C | -0.6804 | -0.60976 | -0.60302 |
| H | -1.37364 | -0.73844 | -1.43578 |
| H | 0.070237 | -1.41443 | -0.6673 |
| C | -1.39337 | -0.71894 | 0.678332 |
| H | -1.1555 | -1.14813 | 1.63829 |

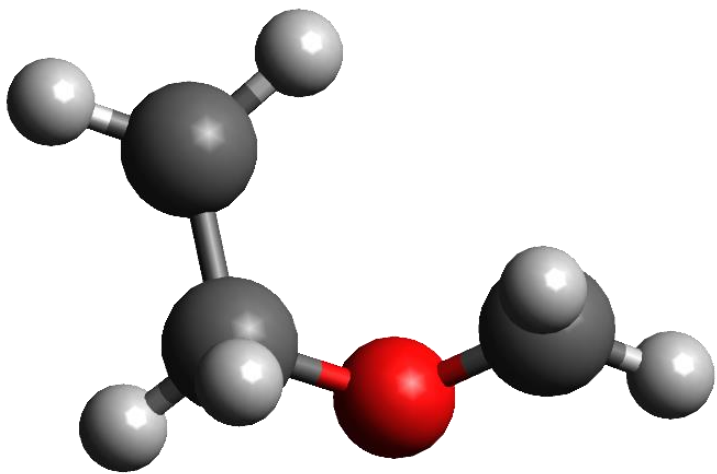

INT6

| | | | |
|---|---|---|---|
| O | 0.002364 | 0.738851 | -0.87612 |
| C | 1.184886 | 1.083496 | -0.32577 |
| H | 1.827686 | 0.290614 | 0.038045 |
| H | -0.37447 | 0.101303 | 1.695535 |
| H | 1.592788 | 2.006234 | -0.70372 |
| C | -0.52202 | -0.50687 | -0.44014 |
| H | -1.3734 | -0.71721 | -1.08671 |
| H | 0.234409 | -1.28406 | -0.63461 |
| C | -0.92349 | -0.49986 | 0.987307 |
| H | -1.64875 | -1.21251 | 1.346183 |

**Transition states**

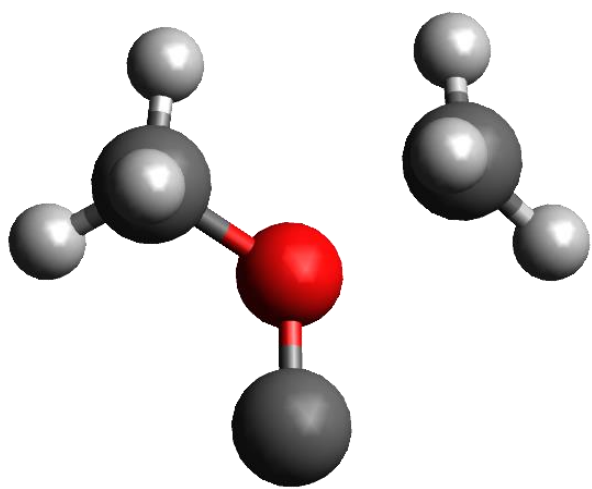

TS1
| | | | |
|---|---|---|---|
| O | -0.43194 | 0.652197 | 0.138771 |
| C | 1.346134 | 0.341825 | 0.125032 |
| H | 1.66682 | 1.371761 | 0.16482 |
| H | 1.47063 | -0.15042 | -0.83015 |
| H | 1.549757 | -0.25407 | 1.004669 |
| C | -1.0735 | -0.65659 | 0.061652 |
| H | -2.13039 | -0.49503 | 0.242225 |
| H | -0.91372 | -1.06891 | -0.9319 |
| H | -0.63119 | -1.2776 | 0.835186 |
| C | -0.85261 | 1.536841 | -0.8103 |

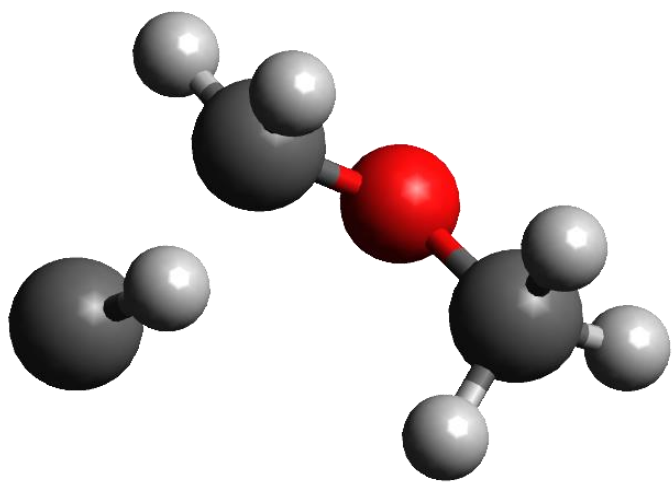

TS2
| | | | |
|---|---|---|---|
| O | -0.47366 | -0.42744 | -0.49836 |
| C | 0.61526 | -0.66692 | 0.260534 |
| H | 1.306684 | -1.34978 | -0.21948 |
| H | 0.388403 | -0.94561 | 1.295384 |
| H | 1.181381 | 0.561942 | 0.814434 |
| C | -1.43246 | 0.413358 | 0.119738 |
| H | -2.27849 | 0.482323 | -0.55593 |
| H | -1.75466 | -0.01019 | 1.074737 |
| H | -1.00396 | 1.405783 | 0.278836 |
| C | 1.844556 | 0.774067 | -0.14369 |

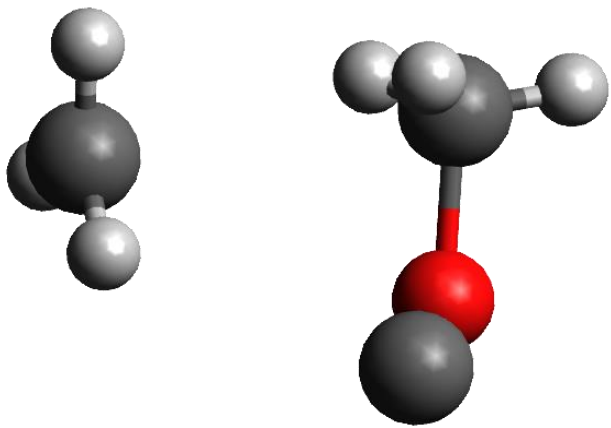

TS3

| | | | |
|---|---|---|---|
| O | -0.39575 | 0.221184 | -0.2275 |
| C | 3.084431 | -0.37305 | -0.16568 |
| H | 2.941806 | 0.581687 | 0.311339 |
| H | 3.075864 | -0.44235 | -1.24055 |
| H | 3.365671 | -1.23004 | 0.423136 |
| C | -0.16581 | -1.4031 | 0.140421 |
| H | -1.17542 | -1.77848 | 0.239528 |
| H | 0.370431 | -1.77904 | -0.72075 |
| H | 0.411258 | -1.3937 | 1.056993 |
| C | 0.013751 | 1.076255 | 0.506387 |

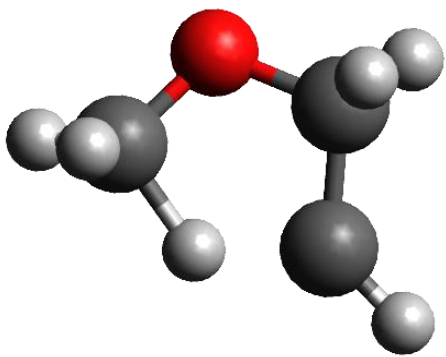

TS4

| | | | |
|---|---|---|---|
| O | 0.20203 | 0.472492 | -1.02909 |
| C | 0.980231 | 0.92593 | 0.020068 |
| H | 1.899862 | 0.3616 | 0.187099 |
| H | 0.185643 | 0.506161 | 1.011347 |
| H | 1.110064 | 2.00249 | -0.00419 |
| C | -0.58395 | -0.64044 | -0.59673 |
| H | -1.50502 | -0.63632 | -1.1833 |
| H | -0.05056 | -1.57843 | -0.8119 |
| C | -0.78636 | -0.45829 | 0.859657 |
| H | -1.45194 | -0.9552 | 1.547035 |

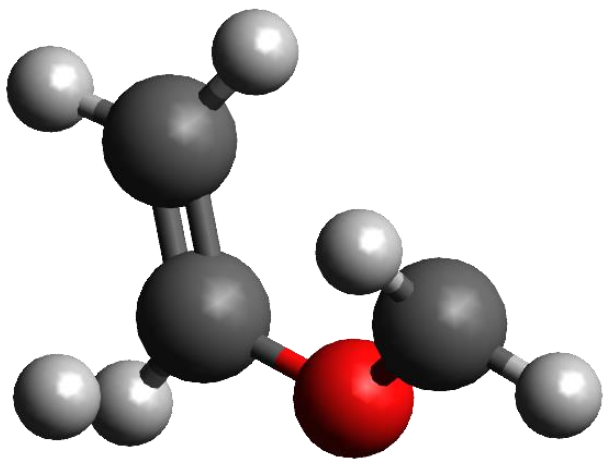

TS5

| | | | |
|---|---|---|---|
| O | 0.016701 | 0.717049 | -0.94624 |
| C | 1.122 | 1.132194 | -0.28903 |
| H | 1.624442 | 0.408603 | 0.339554 |
| H | -0.23769 | 0.062384 | 1.674745 |
| H | 1.655462 | 1.913396 | -0.80248 |
| C | -0.74723 | -0.24852 | -0.36569 |
| H | -1.44365 | -0.66698 | -1.07783 |
| H | 0.401142 | -1.64261 | -0.80326 |
| C | -0.85139 | -0.4622 | 0.958595 |
| H | -1.53979 | -1.21332 | 1.311648 |

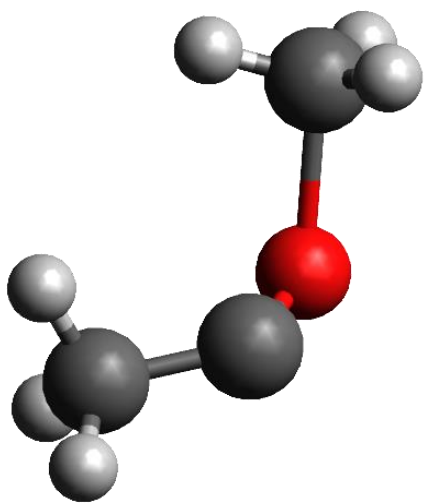

TS6

| | | | |
|---|---|---|---|
| O | -0.87972 | 0.682262 | -0.24887 |
| C | 1.56661 | 0.731278 | -0.2159 |
| H | 2.294426 | 1.361268 | 0.28737 |
| H | 1.641276 | 0.873882 | -1.29676 |
| H | 1.803512 | -0.31354 | 0.011272 |
| C | -1.58832 | -0.95175 | 0.258798 |
| H | -2.46344 | -0.98718 | -0.37146 |
| H | -0.79403 | -1.63912 | 0.005092 |
| H | -1.7636 | -0.79068 | 1.311788 |
| C | 0.183295 | 1.033568 | 0.258667 |

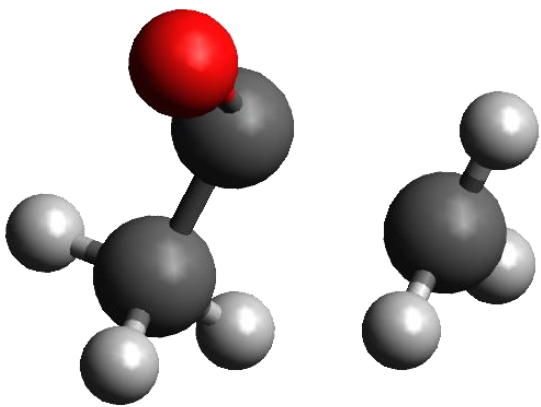

TS7
| | | | |
|---|---|---|---|
| O | 3.23492 | 2.679907 | -0.76783 |
| C | 4.017871 | 0.476388 | -0.14086 |
| H | 5.076618 | 0.700087 | -0.01174 |
| H | 3.8565 | 0.085399 | -1.14763 |
| H | 3.716865 | -0.25183 | 0.604467 |
| C | 3.280057 | 1.798369 | 0.053862 |
| C | 1.353225 | 1.091544 | 0.013159 |
| H | 1.308233 | 0.590138 | -0.94347 |
| H | 0.812076 | 2.020948 | 0.096365 |
| H | 1.388184 | 0.469147 | 0.895935 |

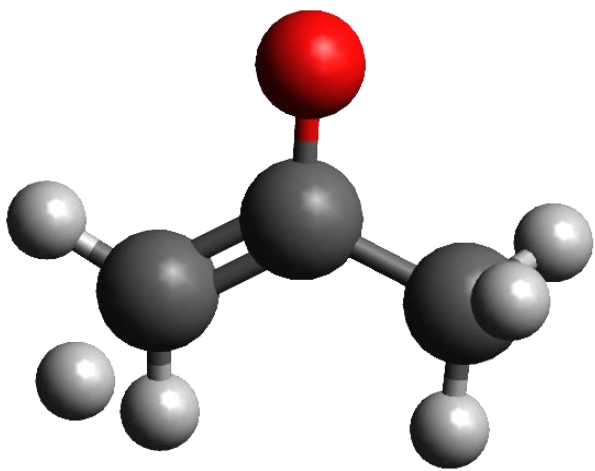

TS8
| | | | |
|---|---|---|---|
| O | -0.57533 | -0.07514 | 1.299034 |
| C | 0.334228 | 1.193075 | -0.42935 |
| H | 0.285126 | 2.099957 | 0.154239 |
| H | -1.2592 | 1.825553 | -1.44391 |
| H | 0.916221 | 1.191042 | -1.33705 |
| C | 0.025478 | -1.31874 | -0.59877 |
| H | 0.593326 | -2.0277 | 0.001587 |
| H | -0.97759 | -1.71887 | -0.7471 |
| H | 0.501173 | -1.18242 | -1.56624 |
| C | -0.05413 | 0.013241 | 0.098047 |

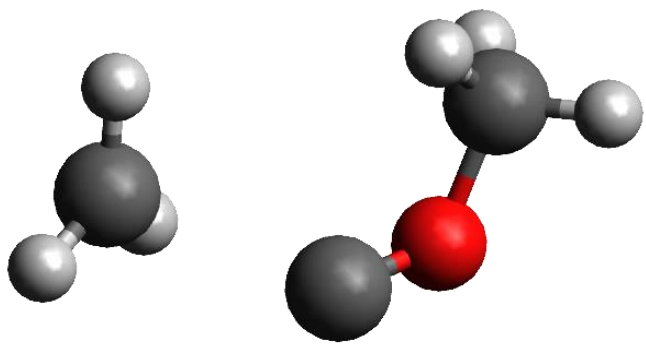

TS9

| | | | |
|---|---|---|---|
| O | -0.04908 | -0.14549 | -0.33582 |
| C | 3.139505 | -0.03853 | -0.15727 |
| H | 3.499054 | 0.885822 | 0.259448 |
| H | 2.973644 | -0.11271 | -1.21892 |
| H | 3.153126 | -0.93182 | 0.445075 |
| C | -0.39772 | -1.47651 | 0.16184 |
| H | -1.48119 | -1.54084 | 0.177343 |
| H | 0.019391 | -2.1946 | -0.53793 |
| H | 0.024385 | -1.59735 | 1.156219 |
| C | 0.645114 | 0.631409 | 0.37334 |

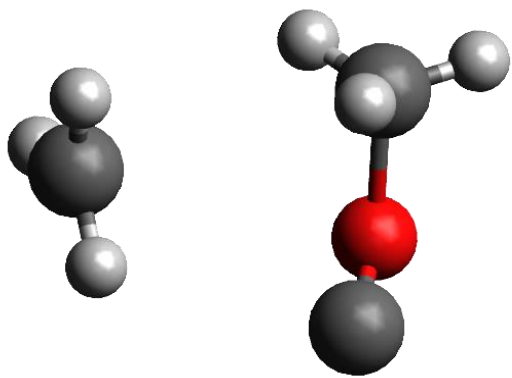

TS10

| | | | |
|---|---|---|---|
| O | -0.83748 | 1.535309 | -0.3447 |
| C | 1.850629 | 0.638001 | -0.29398 |
| H | 2.272172 | 1.470237 | 0.249679 |
| H | 1.717009 | 0.75892 | -1.35855 |
| H | 2.071384 | -0.35124 | 0.074191 |
| C | -0.21532 | 0.862284 | 0.325617 |

**Table S1 Reactants and products (frequencies in cm$^{-1}$)**

| CH$_3$OCH$_3$ | CH$_2$OCCH$_3$ | CH$_3$ | CH$_3$CO | CH$_3$COCH$_2$ | CH$_3$OC | CH$_3$OCCH$_2$ | CO | H$_2$CCHOCH$_2$ | HCCHOCH$_3$ |
|---|---|---|---|---|---|---|---|---|---|
| 236 | 51 | 432 | 107 | 22 | 101 | 112 | 2271 | 223 | 247 |
| 268 | 175 | 1412 | 469 | 369 | 475 | 160 | | 321 | 275 |
| 430 | 238 | 1413 | 867 | 388 | 824 | 320 | | 374 | 314 |
| 984 | 340 | 3144 | 957 | 518 | 1132 | 492 | | 591 | 558 |
| 1135 | 376 | 3321 | 1050 | 533 | 1159 | 677 | | 652 | 649 |
| 1177 | 576 | 3323 | 1354 | 741 | 1381 | 851 | | 725 | 849 |
| 1207 | 901 | | 1461 | 837 | 1483 | 911 | | 886 | 928 |
| 1235 | 913 | | 1463 | 933 | 1490 | 1078 | | 930 | 983 |
| 1278 | 1114 | | 1987 | 1030 | 1528 | 1172 | | 1007 | 1187 |
| 1462 | 1140 | | 3066 | 1070 | 3095 | 1181 | | 1111 | 1187 |
| 1492 | 1194 | | 3165 | 1285 | 3204 | 1258 | | 1212 | 1240 |
| 1495 | 1309 | | 3169 | 1397 | 3209 | 1408 | | 1297 | 1340 |
| 1503 | 1366 | | | 1464 | | 1480 | | 1384 | 1474 |
| 1509 | 1451 | | | 1481 | | 1502 | | 1426 | 1499 |
| 1526 | 1454 | | | 1482 | | 1510 | | 1492 | 1513 |
| 3013 | 1479 | | | 1675 | | 1746 | | 1715 | 1674 |
| 3020 | 3039 | | | 3073 | | 3070 | | 3162 | 3044 |
| 3064 | 3115 | | | 3139 | | 3144 | | 3205 | 3105 |
| 3071 | 3133 | | | 3180 | | 3145 | | 3237 | 3179 |
| 3159 | 3201 | | | 3189 | | 3189 | | 3296 | 3193 |
| 3161 | 3365 | | | 3295 | | 3266 | | 3308 | 3319 |

**Table S2 Intermediates (frequencies in cm$^{-1}$)**

| vdW1 | vdW2 | INT1 | INT2 | INT3 | INT4 | INT5 | INT6 |
|---|---|---|---|---|---|---|---|
| 154 | 93 | 46 | 81 | 71 | 168 | 125 | 92 |
| 240 | 173 | 77 | 140 | 76 | 180 | 203 | 154 |
| 328 | 218 | 102 | 149 | 88 | 315 | 240 | 276 |
| 347 | 270 | 119 | 403 | 100 | 359 | 374 | 378 |
| 387 | 415 | 122 | 454 | 127 | 379 | 543 | 425 |
| 506 | 455 | 162 | 881 | 147 | 806 | 745 | 517 |
| 918 | 992 | 199 | 1021 | 189 | 959 | 942 | 602 |
| 1091 | 1048 | 479 | 1042 | 469 | 964 | 1059 | 836 |
| 1168 | 1148 | 510 | 1094 | 495 | 978 | 1088 | 958 |
| 1207 | 1196 | 825 | 1171 | 864 | 1101 | 1166 | 1065 |
| 1215 | 1235 | 1129 | 1200 | 956 | 1162 | 1185 | 1118 |
| 1279 | 1278 | 1164 | 1351 | 1052 | 1279 | 1222 | 1195 |
| 1455 | 1344 | 1380 | 1390 | 1352 | 1374 | 1307 | 1291 |
| 1475 | 1455 | 1410 | 1465 | 1410 | 1400 | 1374 | 1311 |
| 1489 | 1493 | 1416 | 1472 | 1415 | 1463 | 1464 | 1402 |
| 1494 | 1498 | 1484 | 1489 | 1452 | 1476 | 1481 | 1452 |
| 1497 | 1511 | 1488 | 1500 | 1463 | 1479 | 1498 | 1476 |
| 1513 | 1537 | 1524 | 1504 | 1983 | 1501 | 1515 | 1506 |
| 3069 | 2344 | 3093 | 3009 | 3064 | 3019 | 2963 | 2962 |
| 3074 | 3026 | 3137 | 3059 | 3140 | 3021 | 3019 | 3114 |
| 3156 | 3034 | 3200 | 3060 | 3155 | 3121 | 3078 | 3140 |
| 3157 | 3096 | 3206 | 3134 | 3177 | 3122 | 3095 | 3187 |
| 3203 | 3178 | 3313 | 3154 | 3315 | 3165 | 3161 | 3291 |
| 3206 | 3197 | 3316 | 3176 | 3320 | 3166 | 3264 | 3296 |

**Table S3 Transition states (frequencies in cm⁻¹)**

| TS1 | TS2 | TS3 | TS4 | TS5 | TS6 | TS7 | TS8 | TS9 | TS10 |
|---|---|---|---|---|---|---|---|---|---|
| -1034 | -713 | -845 | -1950 | -873 | -861 | -543 | -1804 | -144 | -415 |
| 179 | 121 | 44 | -289 | 210 | 13 | 16 | 180 | 53 | 48 |
| 222 | 182 | 81 | 191 | 323 | 75 | 200 | 230 | 62 | 272 |
| 303 | 242 | 104 | 381 | 359 | 136 | 253 | 269 | 90 | 512 |
| 342 | 407 | 110 | 619 | 446 | 379 | 353 | 305 | 179 | 556 |
| 444 | 456 | 150 | 706 | 493 | 448 | 436 | 400 | 258 | 891 |
| 783 | 616 | 210 | 851 | 595 | 726 | 608 | 538 | 262 | 1421 |
| 833 | 860 | 279 | 963 | 665 | 765 | 637 | 715 | 487 | 1427 |
| 875 | 997 | 448 | 997 | 698 | 911 | 843 | 876 | 566 | 2120 |
| 982 | 1145 | 523 | 1095 | 845 | 980 | 968 | 885 | 861 | 3124 |
| 1152 | 1185 | 897 | 1107 | 929 | 1049 | 1015 | 981 | 1141 | 3287 |
| 1177 | 1235 | 965 | 1150 | 1049 | 1073 | 1047 | 992 | 1164 | 3297 |
| 1186 | 1270 | 1230 | 1192 | 1112 | 1359 | 1358 | 1064 | 1356 | |
| 1438 | 1293 | 1420 | 1215 | 1213 | 1436 | 1422 | 1232 | 1411 | |
| 1451 | 1461 | 1431 | 1232 | 1283 | 1446 | 1436 | 1371 | 1418 | |
| 1458 | 1493 | 1451 | 1362 | 1379 | 1449 | 1470 | 1392 | 1490 | |
| 1484 | 1499 | 1456 | 1489 | 1428 | 1470 | 1476 | 1469 | 1492 | |
| 1502 | 1508 | 1528 | 1508 | 1484 | 1501 | 1709 | 1487 | 1506 | |
| 3091 | 2400 | 3105 | 1744 | 1624 | 3029 | 3055 | 1572 | | |
| 3110 | 3037 | 3140 | 2980 | 3151 | 3096 | 3116 | 3077 | | |
| 3190 | 3045 | 3254 | 3066 | 3201 | 3124 | 3141 | 3151 | | |
| 3209 | 3113 | 3265 | 3086 | 3233 | 3173 | 3194 | 3183 | | |
| 3269 | 3189 | 3313 | 3192 | 3300 | 3281 | 3276 | 3195 | | |
| 3285 | 3197 | 3320 | 3253 | 3305 | 3292 | 3292 | 3300 | | |

**Table S4 Energies, frequencies and rotational constants along the entrance channel leading to vdW1 formation**

| Energy relative to the C($^3$P) + CH$_3$OCH$_3$ asymptote, (uncorrected for ZPE differences) calculated at the DLPNO-CCSD(T) / AVTZ level / kJ mol$^{-1}$ ||||||||||||||||||||||||
|---|---|---|---|---|---|---|---|---|---|---|---|---|---|---|---|---|---|---|---|---|---|---|---|
| vdW1 |||||||||||||||||||||||C($^3$P) + CH$_3$OCH$_3$ |
| -63.5 | -60.2 | -49.0 | -37.5 | -28.1 | -20.7 | -15.4 | -12.1 | -9.3 | -7.0 | -5.3 | -4.1 | -3.4 | -2.8 | -2.6 | -2.2 | -2.0 | -1.7 | -1.4 | -1.2 | -1.0 | -0.7 |
| **Distance of approach between C($^3$P) and O atom of CH$_3$OCH$_3$ / Å** ||||||||||||||||||||||
| 1.65 | 1.85 | 2.05 | 2.25 | 2.45 | 2.65 | 2.85 | 3.05 | 3.25 | 3.45 | 3.65 | 3.85 | 4.05 | 4.25 | 4.45 | 4.65 | 4.85 | 5.05 | 5.25 | 5.45 | 5.65 | 6.05 |
| **Frequencies calculated at the M06-2X / AVTZ level / cm$^{-1}$** ||||||||||||||||||||||
| 153 | | | | | | | | | | | | | | | | | | | | | |
| 239 | 235 | 171 | 170 | 125 | 75 | 119 | 61 | 36 | 83 | 89 | 80 | 57 | 30 | 19 | 37 | 18 | 35 | 16 | 25 | 11 | 10 |
| 328 | 262 | 236 | 197 | 159 | 116 | 170 | 115 | 82 | 134 | 111 | 101 | 76 | 73 | 25 | 64 | 48 | 54 | 45 | 46 | 41 | 33 |
| 347 | 290 | 248 | 239 | 229 | 204 | 208 | 186 | 215 | 226 | 220 | 219 | 210 | 222 | 219 | 221 | 218 | 216 | 215 | 214 | 213 | 213 |
| 387 | 320 | 311 | 304 | 283 | 259 | 257 | 253 | 267 | 287 | 286 | 280 | 274 | 271 | 270 | 269 | 267 | 266 | 264 | 266 | 264 | 264 |
| 506 | 430 | 425 | 427 | 428 | 425 | 430 | 430 | 430 | 432 | 434 | 434 | 431 | 431 | 430 | 432 | 430 | 430 | 429 | 429 | 428 | 428 |
| 917 | 941 | 954 | 963 | 970 | 973 | 976 | 979 | 978 | 981 | 981 | 981 | 980 | 984 | 983 | 984 | 984 | 984 | 984 | 984 | 984 | 984 |
| 1091 | 1113 | 1125 | 1132 | 1135 | 1135 | 1134 | 1135 | 1139 | 1138 | 1138 | 1137 | 1135 | 1141 | 1141 | 1141 | 1141 | 1141 | 1141 | 1141 | 1141 | 1141 |
| 1168 | 1170 | 1173 | 1172 | 1170 | 1169 | 1170 | 1169 | 1174 | 1175 | 1176 | 1178 | 1173 | 1174 | 1173 | 1175 | 1174 | 1174 | 1174 | 1173 | 1173 | 1173 |
| 1207 | 1212 | 1213 | 1213 | 1208 | 1202 | 1202 | 1201 | 1204 | 1207 | 1207 | 1209 | 1203 | 1204 | 1203 | 1204 | 1203 | 1203 | 1203 | 1203 | 1203 | 1203 |
| 1215 | 1214 | 1220 | 1224 | 1227 | 1228 | 1230 | 1232 | 1233 | 1235 | 1234 | 1234 | 1235 | 1236 | 1237 | 1238 | 1238 | 1238 | 1239 | 1239 | 1240 | 1240 |
| 1279 | 1282 | 1283 | 1283 | 1282 | 1280 | 1280 | 1279 | 1280 | 1280 | 1281 | 1281 | 1280 | 1281 | 1281 | 1281 | 1281 | 1280 | 1280 | 1280 | 1280 | 1280 |
| 1455 | 1460 | 1463 | 1463 | 1463 | 1462 | 1465 | 1462 | 1463 | 1463 | 1463 | 1463 | 1460 | 1464 | 1464 | 1465 | 1464 | 1465 | 1464 | 1464 | 1464 | 1465 |
| 1475 | 1482 | 1487 | 1491 | 1493 | 1490 | 1492 | 1489 | 1492 | 1492 | 1494 | 1494 | 1492 | 1494 | 1494 | 1493 | 1493 | 1494 | 1494 | 1494 | 1494 | 1495 |
| 1489 | 1492 | 1495 | 1497 | 1498 | 1496 | 1496 | 1496 | 1498 | 1495 | 1498 | 1496 | 1494 | 1497 | 1497 | 1498 | 1497 | 1498 | 1497 | 1498 | 1498 | 1498 |
| 1494 | 1496 | 1497 | 1498 | 1500 | 1500 | 1501 | 1501 | 1503 | 1503 | 1503 | 1507 | 1502 | 1504 | 1505 | 1504 | 1504 | 1504 | 1504 | 1505 | 1505 | 1505 |
| 1497 | 1500 | 1507 | 1508 | 1506 | 1503 | 1506 | 1505 | 1504 | 1505 | 1508 | 1508 | 1505 | 1506 | 1506 | 1506 | 1506 | 1506 | 1507 | 1506 | 1507 | 1506 |
| 1513 | 1512 | 1511 | 1512 | 1512 | 1515 | 1520 | 1517 | 1518 | 1519 | 1520 | 1523 | 1513 | 1519 | 1520 | 1521 | 1521 | 1522 | 1522 | 1522 | 1523 | 1522 |
| 3069 | 3058 | 3047 | 3037 | 3029 | 3023 | 3019 | 3016 | 3012 | 3013 | 3012 | 3010 | 3012 | 3007 | 3006 | 3009 | 3005 | 3011 | 3008 | 3011 | 3009 | 3011 |
| 3074 | 3063 | 3052 | 3042 | 3034 | 3029 | 3026 | 3023 | 3019 | 3021 | 3020 | 3019 | 3020 | 3015 | 3012 | 3015 | 3012 | 3018 | 3015 | 3019 | 3016 | 3019 |
| 3156 | 3138 | 3120 | 3104 | 3091 | 3082 | 3076 | 3072 | 3064 | 3069 | 3067 | 3065 | 3065 | 3059 | 3057 | 3059 | 3055 | 3061 | 3058 | 3062 | 3059 | 3063 |
| 3157 | 3139 | 3123 | 3107 | 3095 | 3086 | 3081 | 3076 | 3070 | 3073 | 3072 | 3070 | 3071 | 3065 | 3064 | 3066 | 3062 | 3068 | 3065 | 3069 | 3066 | 3068 |
| 3203 | 3193 | 3183 | 3174 | 3167 | 3164 | 3162 | 3160 | 3157 | 3160 | 3159 | 3158 | 3158 | 3155 | 3155 | 3156 | 3157 | 3156 | 3156 | 3156 | 3157 | 3156 |
| 3206 | 3195 | 3185 | 3176 | 3168 | 3165 | 3163 | 3161 | 3158 | 3162 | 3162 | 3162 | 3162 | 3157 | 3157 | 3157 | 3158 | 3157 | 3158 | 3157 | 3158 | 3157 |
| **Rotational constants calculated at the M06-2X / AVTZ level / cm$^{-1}$** ||||||||||||||||||||||
| A | 0.32 | 0.31 | 0.31 | 0.31 | 0.31 | 0.30 | 0.30 | 0.30 | 0.30 | 0.30 | 0.30 | 0.30 | 0.30 | 0.31 | 0.31 | 0.31 | 0.31 | 0.32 | 0.31 | 0.32 | 0.32 | 0.32 |
| B+C | 0.24 | 0.23 | 0.21 | 0.19 | 0.17 | 0.16 | 0.15 | 0.14 | 0.13 | 0.12 | 0.11 | 0.11 | 0.10 | 0.09 | 0.08 | 0.08 | 0.07 | 0.07 | 0.06 | 0.06 | 0.06 | 0.05 |

**Table S5 Reactions involved in CH$_3$OCH$_3$ production and destruction**

| | Reaction | ΔE kJ/mol | α | β | γ | F$_0$ | g | ref |
|---|---|---|---|---|---|---|---|---|
| 1. | CH$_3$ + CH$_3$O → CH$_4$ + H$_2$CO<br>→ CH$_3$OCH$_3$ | -348<br>-328 | 4.0e-11<br>1.37e-12 | 0<br>-0.96 | 0<br>0 | 1.6<br>10 | 0<br>0 | (Tsang & Hampson 1986b),<br>(Tennis *et al.* 2021) |
| 2. | CH$_3$ + CH$_3$OH$_2^+$ → CH$_3$OHCH$_3^+$ + H<br>→ CH$_3$CH$_2$OH$_2^+$ + H | +45<br>+12 | | | | | | |
| 3. | CH$_3$OH$_2^+$ + CH$_3$OH → CH$_3$OHCH$_3^+$ + H$_2$O | -59 | 1.1e-10 | -1.0 | 0 | 1.60 | 0 | (Anicich 2003) |
| 4. | CH$_3^+$ + CH$_3$OH → CH$_3$OHCH$_3^+$ + hν<br>→ CH$_4$ + H$_2$COH$^+$ | -335<br>-249 | 7.8e-12<br>2.3e-9 | -1.1<br>-0.5 | 0<br>0 | 10<br>1.8 | 0<br>0 | (Jarrold *et al.* 1986, Herbst 1987)<br>(Anicich 2003) |
| 5. | NH$_3$ + CH$_3$OHCH$_3^+$ → CH$_3$OCH$_3$ + NH$_4^+$ | -58 | 1.0e-9 | 0 | 0 | 1.8 | 0 | (Skouteris *et al.* 2019) |
| 6. | CH$_3$OHCH$_3^+$ + e$^-$ → CH$_3$OCH$_3$ + H<br>→ CH$_3$OH + CH$_3$<br>→ CH$_4$ + O + CH$_3$ | -529<br>-610<br>-235 | 8.5e-8<br>9.2e-7<br>7.5e-7 | -0.7<br>-0.7<br>-0.7 | 0<br>0<br>0 | 1.6<br>1.6<br>1.6 | 0<br>0<br>0 | (Hamberg *et al.* 2010) |
| 7. | H$^+$ + CH$_3$OCH$_3$ → CH$_3$OCH$_3^+$ + H<br>→ CH$_4$ + H$_2$COH$^+$ | -356<br>-693 | 2.5e-9<br>2.5e-9 | -0.5<br>-0.5 | 0<br>0 | 3<br>3 | 0<br>0 | By comparison with H$^+$ + H$_2$CO, CH$_3$OH but without introducing CH$_3$OCH$_2^+$. |
| 8. | He$^+$ + CH$_3$OCH$_3$ → CH$_2^+$ + CH$_3$O + H + He<br>→ CH$_3^+$ + H + H$_2$CO + He<br>→ HCO$^+$ + CH$_3$ + H$_2$ + He | -585<br>-1082<br>-1205 | 7e-11<br>4e-10<br>6e-10 | 0<br>0<br>0 | 0<br>0<br>0 | 1.6<br>1.6<br>1.6 | 0<br>0<br>0 | (Ascenzi *et al.* 2019) |
| 9. | H$_3^+$ + CH$_3$OCH$_3$ → CH$_3$OHCH$_3^+$ + H$_2$<br>→ CH$_3$OH + CH$_3^+$ + H$_2$<br>→ CH$_4$ + CH$_3^+$ + H$_2$O<br>→ CH$_3$OH + CH$_5^+$<br>→ H$_2$COH$^+$ + CH$_4$ + H$_2$<br>→ CH$_3$OCH$_2^+$ + H$_2$ + H$_2$<br>→ C$_2$H$_5^+$ + H$_2$O + H$_2$ | -354<br>-18<br>-126<br>-195<br>-267<br>-236<br>-244 | 5.6e-10<br>4.2e-10<br>9.4e-10<br>3.8e-10<br>1.2e-9<br>0(7.1e-10)<br>4.7e-10 | -0.5 | 0 | 1.8 | 0 | (Lee *et al.* 1992)<br>CH$_3$OCH$_2^+$ not included in our network. |
| 10. | C$^+$ + CH$_3$OCH$_3$ → C + CH$_3$OCH$_3^+$<br>→ CH$_3$ + CH$_3$CO$^+$ →<br>CH$_3^+$ + CH$_3$CO | -140<br>-806<br>-537 | 2.0e-9<br>2.0e-9<br>0 | -0.5<br>-0.5 | 0<br>0 | 2<br>2 | 0<br>0 | Rate constant and products guessed by comparison with C$^+$ + CH$_3$OH, (CH$_3$)$_2$CO (Anicich 2003). |
| 11. | C + CH$_3$OCH$_3$ → CH$_3$ + CH$_3$ + CO | -345 | 1.3e-11 | -1.0 | 0 | 1.4 | 8 | 50-300K, this work (we do not consider CH$_3$OC formation included into CH$_3$ + CH$_3$ + CO channel). We use k=1.3e-10 cm$^3$.s$^{-1}$ in the 10-49K range. |
| 12. | OH + CH$_3$OCH$_3$ → CH$_3$OCH$_2$ + H$_2$O | -94 | 1.0e-11 | 0 | 0 | 2 | 0 | Around 10-50K only. (Shannon *et al.* 2014a, Klippenstein 2017) |
| 13. | $HCO^+ + CH_3OCH_3 \rightarrow CH_3OHCH_3^+ + CO$ | | | 2.1e-9 | -0.5 | 0 | 1.4 | 0 | (Anicich 2003) |
| 14. | $S^+ + CH_3OCH_3 \rightarrow S + CH_3OCH_3^+$ | -54 | 1.0e-9 | -0.5 | 0 | 1.6 | 0 | (Decker *et al.* 2000) with simplified products. |
| 15. | $Cl + CH_3OCH_3 \rightarrow HCl + CH_3OCH_2$ | -31 | 1.80e-10 | 0 | 0 | 1.4 | 0 | (Jenkin *et al.* 2010) |

| | Reaction | ΔE kJ/mol | Branching ratio | γ (K) | Ref |
|---|---|---|---|---|---|
| 16. | $C + s\text{-}CH_3OH \rightarrow s\text{-}CH_3OCH$<br>$\rightarrow s\text{-}CH_4 + s\text{-}CO$<br>$\rightarrow s\text{-}CH_3CHO$ | -396<br>-691<br>-672 | 0.2<br>0.5<br>0.3 | 0<br>0<br>0 | This reaction have been studied in the gas phase being barrierless and leading mostly to $CH_3$ + HCO through $CH_3OCH$ intermediate formation (Shannon *et al.* 2014b). On Ice some $CH_3OCH$ will be stabilized and $CH_3$ + HCO should give $CH_3CHO$ and $CH_4$ + CO as $CH_3$ and HCO will stay close on Ice. This reaction may also produce some $C_2H_3OH$ and $c\text{-}C_2H_4O$. |
| 17. | $s\text{-}H + s\text{-}CH_3OCH \rightarrow s\text{-}CH_3OCH_2$<br>$\rightarrow s\text{-}C_2H_5O$<br>$\rightarrow s\text{-}CH_3 + s\text{-}H_2CO$ | -334<br>-346<br>-288 | 0.5<br>0.5<br>0 | 0<br>0 | The TS for $CH_3OCH_2 \rightarrow C_2H_5O$ is located 207 kJ/mol above the $CH_3OCH_2$ energy, so -126 kJ/mol below the entrance level. Some $CH_3OCH_2$ may dissociate into $CH_3$ + $H_2CO$. |
| 18. | $s\text{-}H + s\text{-}CH_3OCH_2 \rightarrow s\text{-}CH_3OCH_3$<br>$\rightarrow s\text{-}CH_4 + s\text{-}H_2CO$ | -401<br>-383 | 1<br>0 | 0 | The TS for $CH_3OCH_3 \rightarrow CH_4 + H_2CO$ is located 479 kJ/mol above the $CH_3OCH_3$ energy, so $CH_3OCH_3$ cannot dissociate. |
| 19. | $s\text{-}HCO + s\text{-}CH_3OCH_2 \rightarrow s\text{-}CH_3OCH_3 + s\text{-}CO$ | -332 | 1 | 0 | We neglect $CH_3OCH_2CHO$ formation which is however likely not negligible. |
| 20. | $s\text{-}H + s\text{-}CH_3OCH_3 \rightarrow s\text{-}CH_3OCH_2 + s\text{-}H_2$<br>$\rightarrow s\text{-}CH_3 + s\text{-}CH_3OH$ | -25<br>-98 | 1<br>0 | 4040 (1550i)<br>? | This work (M06-2X/AVTZ level) in good agreement with (Takahashi *et al.* 2007) who obtain a barrier height equal to 4090 K. |
| 21. | $s\text{-}CH_3O + s\text{-}CH_3OCH_3 \rightarrow s\text{-}CH_3OCH_2 + s\text{-}CH_3OH$ | -35 | 1 | 1900(1200i) | This work (M06-2X/AVTZ level) |
| 22. | $s\text{-}CH_2OH + s\text{-}CH_3OCH_3 \rightarrow s\text{-}CH_3OCH_2 + s\text{-}CH_3OH$ | 0 | 0 | 7900(1730i) | This work (M06-2X/AVTZ level) |
| 23. | $s\text{-}CH_3 + s\text{-}CH_3O \rightarrow s\text{-}CH_3OCH_3$<br>$\rightarrow s\text{-}CH_4 + s\text{-}H_2CO$ | -345<br>-336 | 0.9<br>0.1 | 0<br>0 | Gas phase branching ratio of (Tsang & Hampson 1986a) favor H atom abstraction but not for the study of (Enrique-Romero *et al.* 2022) on Ice reactions where $CH_3O$ and $CH_2OH$ are not good H donor. The TS for $CH_3OCH_3 \rightarrow CH_4 + H_2CO$ is located 479 kJ/mol above the $CH_3OCH_3$ energy, so $CH_3OCH_3$ cannot dissociate. |